\documentclass[acmsmall]{acmart}

\AtBeginDocument{%
  }

\setcopyright{rightsretained}
\acmDOI{10.1145/3660782}
\acmYear{2024}
\copyrightyear{2024}
\acmSubmissionID{fse24main-p159-p}
\acmJournal{PACMSE}
\acmVolume{1}
\acmNumber{FSE}
\acmArticle{75}
\acmMonth{7}
\received{2023-09-28}
\received[accepted]{2024-04-16}

\usepackage{multirow}
\usepackage{todonotes}
\usepackage{graphicx}
\usepackage{float}
\usepackage{subfigure}
\usepackage{makecell}
\usepackage{tcolorbox}

\begin{document}

\title[\textsc{Epitome}]{Enhancing Function Name Prediction using Votes-Based Name Tokenization and Multi-Task Learning}

\author{Xiaoling Zhang}
\orcid{0009-0001-0906-1817}
\affiliation{%
  \institution{Beijing Key Laboratory of IOT Information Security Technology, Institute of Information Engineering, CAS; School of Cyber Security, University of Chinese Academy of Sciences}
  \city{Beijing}
  \country{China}}
\email{zhangxiaoling@iie.ac.cn}

\author{Zhengzi Xu}
\authornote{Zhiqiang Shi and Zhengzi Xu are the corresponding authors.}
\orcid{0000-0002-8390-7518}
\affiliation{%
  \institution{School of Computer Science and Engineering, Nanyang Technological University}
  \country{Singapore}}
\email{zhengzi.xu@ntu.edu.sg}

\author{Shouguo Yang}
\orcid{0000-0003-4385-8261}
\email{yangshouguo@iie.ac.cn}
\affiliation{%
    \institution{Beijing Key Laboratory of IOT Information Security Technology, Institute of Information Engineering, CAS; School of Cyber Security, University of Chinese Academy of Sciences}
  \city{Beijing}
  \country{China}}
\author{Zhi Li}
\orcid{0000-0001-7071-2976}
\email{lizhi@iie.ac.cn}
\affiliation{%
   \institution{Beijing Key Laboratory of IOT Information Security Technology, Institute of Information Engineering, CAS; School of Cyber Security, University of Chinese Academy of Sciences}
  \city{Beijing}
  \country{China}}
  
\author{Zhiqiang Shi}
\authornotemark[1]
\orcid{0000-0001-6168-8003}
\email{shizhiqiang@iie.ac.cn}
\affiliation{%
    \institution{Beijing Key Laboratory of IOT Information Security Technology, Institute of Information Engineering, CAS; School of Cyber Security, University of Chinese Academy of Sciences}
  \city{Beijing}
  \country{China}}
  
\author{Limin Sun}
\orcid{0000-0003-2745-7521}
\email{sunlimin@iie.ac.cn}
\affiliation{%
    \institution{Beijing Key Laboratory of IOT Information Security Technology, Institute of Information Engineering, CAS; School of Cyber Security, University of Chinese Academy of Sciences}
  \city{Beijing}
  \country{China}}
\renewcommand{\shortauthors}{Zhang and Xu, et al.}

\begin{abstract}
  Reverse engineers would acquire valuable insights from descriptive function names, which are absent in publicly released binaries. 
  Recent advances in binary function name prediction using data-driven machine learning show promise.
  However, existing approaches encounter difficulties in capturing function semantics in diverse optimized binaries and fail to reserve the meaning of labels in function names. 
  We propose \textsc{Epitome}, a framework that enhances function name prediction using votes-based name tokenization and multi-task learning, specifically tailored for different compilation optimization binaries. 
  \textsc{Epitome} learns comprehensive function semantics by pre-trained assembly language model and graph neural network, incorporating function semantics similarity prediction task, to maximize the similarity of function semantics in the context of different compilation optimization levels. 
  In addition, we present two data preprocessing methods to improve the comprehensibility of function names.  
  We evaluate the performance of \textsc{Epitome} using 2,597,346 functions extracted from binaries compiled with 5 optimizations (O0-Os) for 4 architectures (x64, x86, ARM, and MIPS). 
  \textsc{Epitome} outperforms the state-of-the-art function name prediction tool by up to 44.34\%, 64.16\%, and 54.44\% in precision, recall, and F1 score, while also exhibiting superior generalizability.
\end{abstract}




\maketitle
\section{INTRODUCTION}

Reverse engineering is a complex process essential to software-security tasks such as vulnerability discovery, malware analysis, and software debugging~\cite{8418614,7961513}. 
APIs and function names often are used as beacons to help reverse engineers comprehend the behavior of the program's components~\cite{Votipka,b12}.
The loss of valuable information such as file names, function names, and compilation optimization information in stripped binaries, poses a challenge in tracing the binary back to its source-level details. 
Consequently, when confronted with a contiguous section of executable code, attaining comprehension of its functionality can be challenging and time-consuming~\cite{7546501}.

Disassemblers play a pivotal role in the analysis of binaries. Assembly language, derived from machine code using disassemblers, often lacks high-level information, which can make it a challenging and intricate task to read and comprehend.
In an effort to enhance the legibility of binary code, reverse engineers have been exploring the realm of decompilation. This process involves elevating assembly code to a more human-readable format, resembling C-like pseudocode. 
However, an inherent challenge in the decompilation process is the irreversible loss of certain information that occurs during compilation. As a result, decompilers often assign address-related placeholders as default function names. This practice significantly hinders the comprehension of the program's logic and structure. Consequently, despite the availability of disassemblers and advanced decompilation tools, reverse engineering remains a labor-intensive endeavor.
The recovery of original function names holds immense importance in this context, as these names serve as valuable indicators that help infer the purpose and functionality of each function in the code~\cite{Votipka}.

Recent advancements in the field~\cite{b6,b10, b44,b7} have illuminated the potential of leveraging machine learning models for the prediction of function names. These models exhibit the remarkable ability to autonomously acquire valuable semantic insights from binary, contributing to the process of function name recovery.
However, a complex landscape of challenges arises from the inherent variability in compilation optimizations and hardware architectures. This diversity manifests even when the source code is identical, adding a layer of intricacy to the task at hand. Notably, existing methodologies encounter notable difficulties when confronted with binaries compiled under different optimization levels and on diverse hardware architectures~\cite{b44, b6}. In cases where the compilation optimization level of binaries is not distinguished, the model's performance dips significantly, yielding a precision as low as approximately 40\%~\cite{b6}.
The performance of established methods~\cite{b44} exhibits variations contingent upon the optimization levels employed during compilation. 
When predicting function name on the x64 architecture, a significant 60.2\% decrease in F1 score can be observed between the O3 and O0 optimization levels.
The coexistence of binaries compiled under different optimization levels within the dataset appears to obstruct the effective learning of discerning patterns by prior methodologies, further impeding their overall performance.
In conclusion, accurately predicting function names for binaries lacking optimization level information remains a formidable challenge in the field, one that warrants continued research.

The diversity of function names, influenced by a multitude of naming conventions and personal preferences, presents a considerable challenge in the realm of name prediction. 
For instance, research by Feitelson et al.~\cite{feitelson2020developers} has shed light on the fact that different developers often assign distinct names to the same functions, resulting in a mere 6.9\% probability of selecting an identical name. 
This significant variance in naming choices among developers also gives rise to out-of-vocabulary (OOV) problems~\cite{bird2009natural}. 
These problems occur when a word within a function name does not exist in the established name vocabulary.
The amalgamation of such a diverse range of function names and the persistent OOV problem compounds the complexity of the learning process for models endeavoring to establish a meaningful relationship between assembly code and function names.
Given these fundamental observations, the careful handling of the function naming process becomes paramount. 
This, in turn, can significantly enhance the efficacy of deep learning-based methods aimed at predicting function names.

In this research paper, we introduce \textsc{Epitome}, a comprehensive multi-task learning framework specifically crafted to address the intricate challenge of predicting function names for binaries compiled under varying optimization levels. \textsc{Epitome} encompasses two pivotal tasks: the function name prediction task and the function semantics similarity prediction task.
By incorporating the function semantics similarity prediction task, our framework acquires semantic knowledge that significantly augments its ability to comprehend code generated from a wide spectrum of compilation optimizations. Furthermore, to mitigate the issues stemming from the diversity of function names and OOV problems, we propose a novel approach: a votes-based function name tokenization method. This method involves segmenting function names into meaningful labels.
Extensive experimentation demonstrates the prowess of \textsc{Epitome}, achieving remarkable results with a precision of 73.13\%, a recall of 70.84\%, and an F1 score of 71.96\% across a diverse set of hardware architectures and optimization levels. These results outshine the current state-of-the-art function name prediction technique by a substantial margin, surpassing it by up to 54.44\% in terms of F1 score.
\textsc{Epitome} also exhibits exceptional generalizability when assessed across two unique sets of unseen binaries. 
The first evaluation dataset comprises binaries that share thematic similarities with the domains represented in our dataset, while the second evaluation dataset includes binaries that diverge from the domain-specific knowledge contained within our dataset. 
\textsc{Epitome} exceeds the capabilities of state-of-the-art methods, delivering a 45.59\% increase in accuracy on the domain-knowledge shared evaluation dataset. For the domain-knowledge deviated evaluation dataset, \textsc{Epitome} continues to outshine existing approaches, achieving a 61.19\% boost in accuracy. Furthermore, our ablation studies provide compelling evidence of the efficacy of \textsc{Epitome}'s design, particularly in its incorporation of the function semantics similarity prediction task. For instance, this multi-task learning framework can enhance the F1 score of \textsc{Epitome} by a notable 5.97\%.
The votes-based function name tokenization method not only reduces the occurrence of OOV words but also preserves valuable semantic information. For instance, in the context of x86, a remarkable 94.61\% of OOV words are effectively mitigated, and Epitome’s F1 score is improved by 5.6\%.
When applied to real-world IoT firmware, \textsc{Epitome} showcases its potential in predicting meaningful and useful function names, marking a substantial advancement in the field.

The contributions of this paper can be summarized as follows:
\begin{itemize}
    \item  We present \textsc{Epitome}, a multi-task learning framework designed to tackle the intricate task of predicting function names within a wide range of optimized binaries. \textsc{Epitome} leverages comprehensive function semantics learning across different compilation optimization levels through a dedicated function semantics similarity prediction training task.
    \item To address the diversity of function names and mitigate the OOV challenge, we introduce a novel votes-based function name tokenization method. This approach effectively segments function names into meaningful tokens, enhancing the overall robustness of the framework.
    \item Extensive experiments are conducted to assess the performance of \textsc{Epitome}. The results demonstrate that \textsc{Epitome} surpasses the current state-of-the-art in function name prediction. Furthermore, these experiments underscore the framework's exceptional generalizability, the effectiveness of its components, and its practical applicability. 
\end{itemize}

\section{Challenges AND MOTIVATION}
\subsection{Challenges} \label{challange}
To predict function names, it is essential to model the semantics of binary functions compiled with various optimization levels, process their corresponding names, and learn the valuable patterns between semantics and names. In this context, we tackle three challenges associated with this task.

\begin{figure}[bt]
    \vspace{-10pt}
    \centering
    \setlength{\abovecaptionskip}{-2pt}
    \subfigure[O1.]{
        \includegraphics[width=0.32\linewidth]{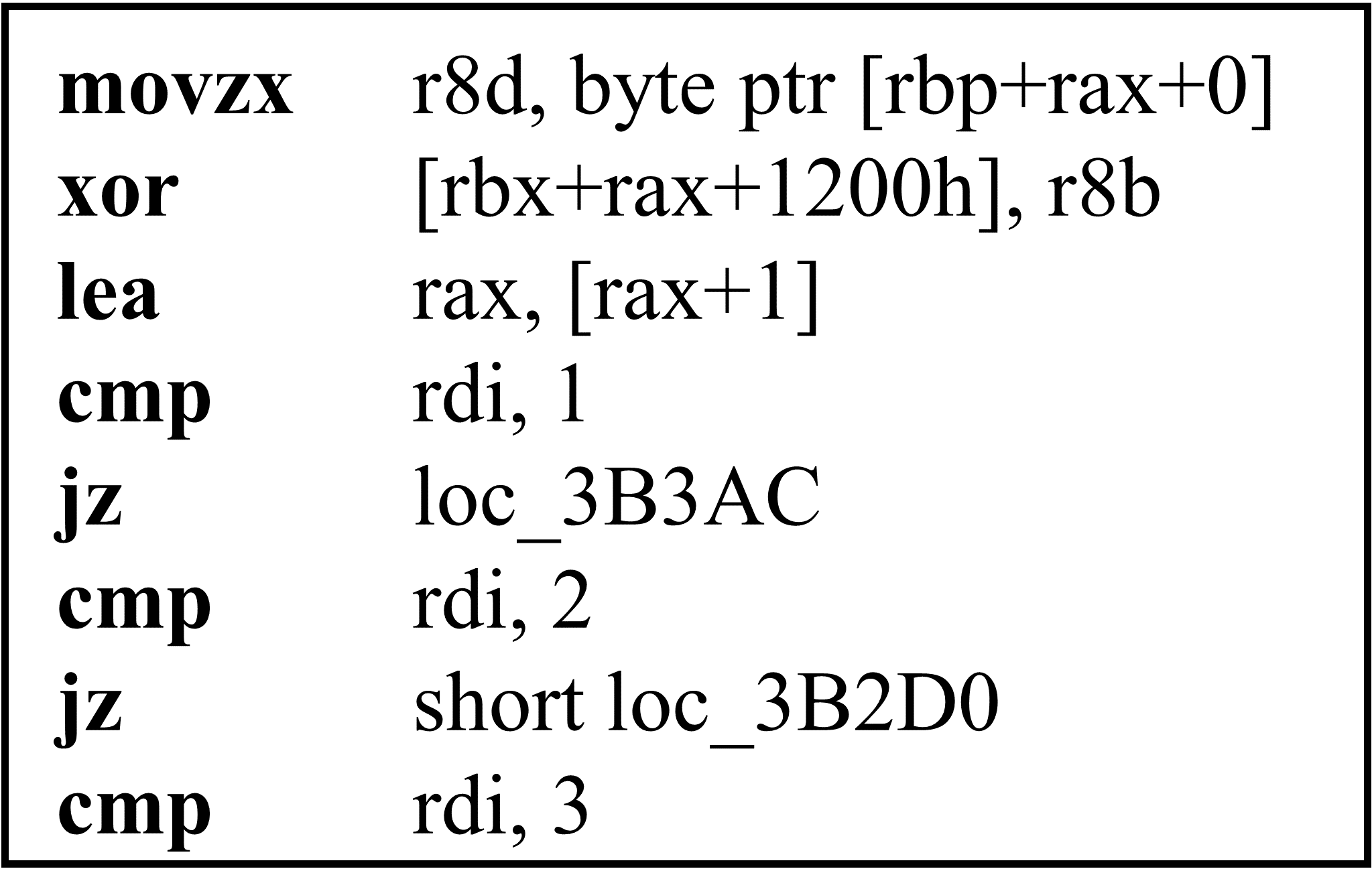}}
    \quad
    \quad
    \centering
    \subfigure[O3.]{
        \includegraphics[width=0.4\linewidth]{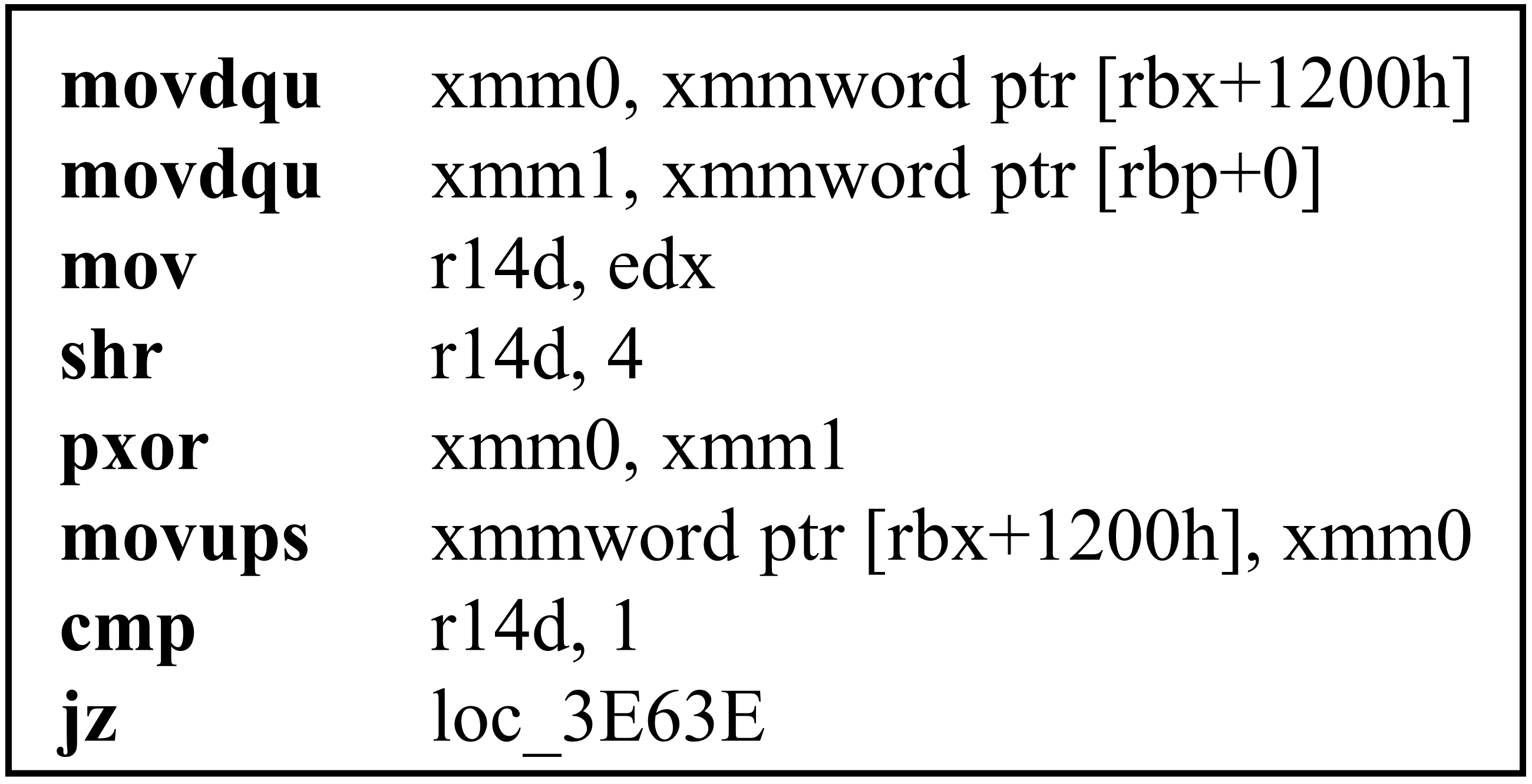}}
    \caption{Code Snippet of Function ocb\_encrypt at Different Compilation Optimization Levels.}
   \label{fig:example}
   \vspace{-10pt}
\end{figure}

\noindent \textbf{C1:\ Limited Semantic Information.} \label{C1}
Different from source code which contains rich high-level symbol information such as variable names, and function names, stripped binaries lack descriptive information, resulting in extremely limited information in assembly code.
This limited information poses challenges for tasks like function name prediction. 
According to the statistics of~\cite{Xu2019Source, fernandes2018structured}, roughly 33\% of tokens in names can be directly copied from tokens in the source code of function bodies by leveraging copy mechanism~\cite{gu2016incorporating}. 
In contrast, the names and types of identifier and function parameters,  which carry valuable information that can serve as strong hints for function name prediction, are discarded in stripped binaries~\cite{host2009debugging, li2021context}.

\noindent \textbf{C2:\ Variety of Binary Code Under Diverse Compilation Optimization Levels.} \label{C2}
Diverse compilation optimization levels~\cite{du2022automatic} can introduce substantial variations in the characteristics of binary code, even when originating from the same source code, as highlighted in previous studies~\cite{ding2019asm2vec, b40}. 
To illustrate the challenges posed by these optimization levels, consider the function \textit{ocb\_encrypt}. The composition and structure of instructions within this function diverge significantly depending on the chosen optimization level. At O3, it encompasses more than 400 instructions and features over 80 basic blocks, while at O1, it consists of approximately 200 instructions and around 40 basic blocks. Figure~\ref{fig:example} provides a visual representation of this divergence in code snippets.
The utilization of various optimization levels presents formidable obstacles for models aspiring to grasp the semantic essence of functions. Moreover, in real-world scenarios, the necessity to handle an array of optimized binaries arises, often compounded by the fact that information regarding compilation optimization is discarded during the stripping process. This inherent limitation further obstructs the model's ability to decipher the intricacies of binaries.

\noindent \textbf{C3:\ Various Naming Methods.} \label{C3}
Function names often involve word abbreviations, domain-specific jargon, and specific programming language conventions~\cite{hindle2016naturalness, scanniello2017fixing}. 
Different developers can use completely different words to refer to the same concepts and operations, and have the flexibility to choose from various naming conventions that are suggested by research~\cite{beck2007implementation}, leading to a diversity of function names, as shown in Table~\ref{tab:name_method}. 
The common naming conventions use capital letters, underscores, and numbers as delimiters to separate words. 
However, developers often omit these delimiters, resulting in function names that appear as continuous strings of characters without clear word boundaries.
The diversity of function names leads to label sparsity, meaning that the same name semantics can have different forms. This results in a large vocabulary of labels, making it challenging to accurately predict the correct label.
In addition, incorrect function name tokenization can result in ambiguous words. 
Research~\cite{avidan2017effects} has shown that names with misleading tokens are worse than meaningless names (like consecutive letters of the alphabet) because they can mislead reverse engineers during analysis.
Variability in naming conventions exacerbates the label sparse issue and OOV problem, which negatively impacts the performance of the model.

\begin{table}[tb]
    \centering
    \vspace{-8pt}
    \caption{Categories of Function Name Methods}
    \begin{tabular}{p{3cm}p{4cm}p{4.5cm}}
    \toprule
        {}&\textbf{Category}&\textbf{Example} \\
        \midrule
        \multirow{4}{*}{Common Name} & camelCase & getTableSize, copyRawBlock \\
         &PascalCase & DisplayInfo, FreeSigner\\
         &snake\_case& set\_rand, bin\_set \\ 
         &Number delimiter& x2realloc, name2oid \\ 
         \hline
         \multirow{3}{*}{Uncommon Name}
         &Word concat&getoutput, settags \\
         &Abbreviation concat&qualdev, incpos \\ 
         &Abbreviation\&Word concat& showmsg, hashcmd \\ 
         \bottomrule
    \end{tabular}
    \vspace{-6pt}
    \label{tab:name_method}
\end{table}

\subsection{Existing Techniques and Our Motivations} \label{relate_work}
Previous studies focus on utilizing deep learning techniques to predict names for binary functions. Regrettably, their efforts were inadequate in adequately addressing the aforementioned challenges, thereby directly motivating our proposition of \textsc{Epitome}.

NFRE~\cite{b6} proposes a structure-sensitive instruction embedding method to generate names for functions. 
Different compilation optimization levels have a significant impact on the semantics of functions generated by NFRE, making it difficult to address the \textbf{C2}.
In addition, the instruction representations used in NFRE lack internal semantics, which hinders the proper addressing of \textbf{C1}. 
NFRE falls short in handling irregular naming conventions and lacks the ability to identify software domain-specific morphological words, thus neglecting the \textbf{C3}. 

SymLM~\cite{b44} learns the execution behavior of the calling context and function instructions to predict the function name.
It is necessary for SymLM to access the optimization level before predicting, but this information is lost in stripped binaries.
Moreover, the performance of SymLM varies greatly under different optimization levels.
It appears that SymLM has not adequately addressed \textbf{C2}.
The tokenization method proposed by SymLM may impede the comprehension of individual labels.
As illustrated, SymLM mistakenly tokenizes `fork' into `for' and `k'. The words `for' have unique meanings different from `fork', which will mislead reverse analysts, as shown in Table~\ref{tab:oov}.
Therefore, it can be concluded that SymLM does not effectively address \textbf{C3}.

\section{Overview}
\begin{figure}[tb]
    \centering
    \vspace{-8pt}
    \includegraphics[scale=0.4]{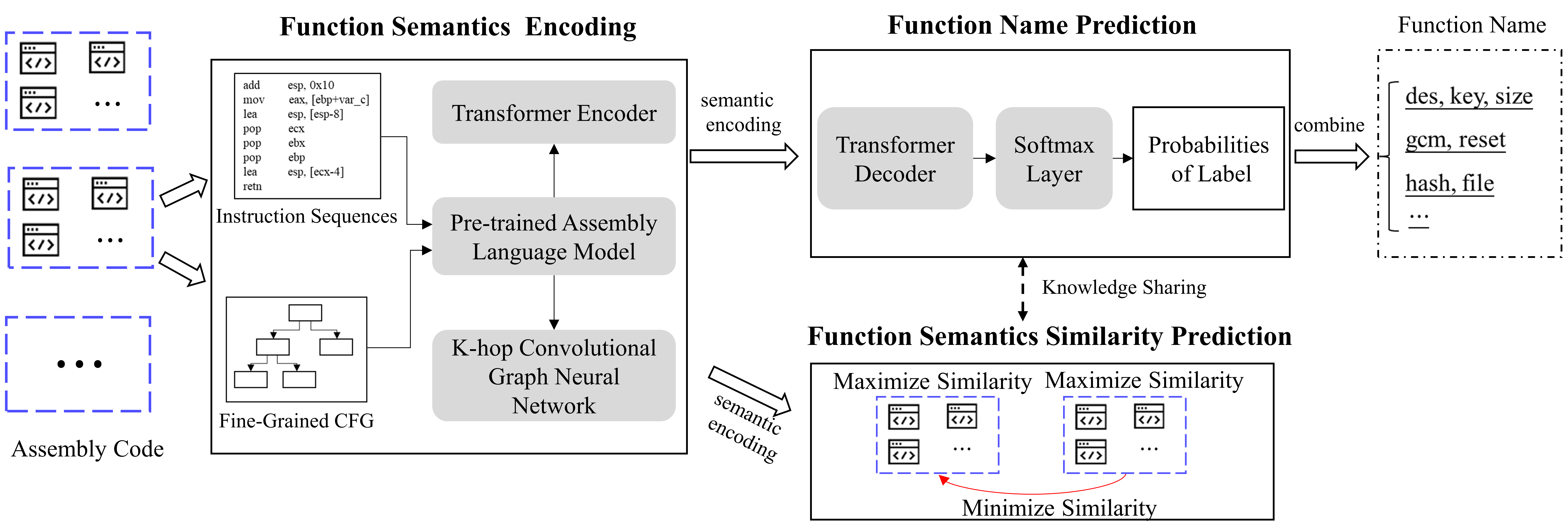}
    \caption{The Overall Workflow of \textsc{Epitome}. }
    \Description{The Overall Workflow of \textsc{Epitome}.The Goal of Functional Semantics Similarity Prediction Is to Maximize the Similarity of the Functions in the Same Group (Blue Dashed Box) And Minimize the Similarity of Functions Across Different Groups. }
    \label{fig:arch}
    \vspace{-8pt}
\end{figure}

To address the challenges summarized in Section~\ref{challange}, we propose \textsc{Epitome}, a multi-task learning framework designed specifically for function name prediction in binaries compiled with different optimization levels.
\textsc{Epitome} consists of two tasks: function name prediction and function semantics similarity prediction.
The multi-task learning framework leverages the shared knowledge between the two tasks, to enhance the generalization and performance of our model.
Figure~\ref{fig:arch} presents \textsc{Epitome}’s workflow. 
The input of \textsc{Epitome} is functions compiled with different optimization levels from the same source code (i.e., the functions in the blue dashed box). Functions in different blue dashed boxes have different function names.
\textsc{Epitome} extracts assembly instructions and control flow information of functions, which are leveraged by \textbf{Function Semantics Encoding} module to learn the comprehensive function semantics. 
The semantic encoding generated by this module is then used to predict function names, as well as to calculate the similarity between input functions. 
\textbf{Function Name Prediction} module learns the mapping between comprehensive function semantics embedding and their corresponding names, and the final output is the most suitable names.
\textbf{Function Semantics Similarity Prediction} module calculates similarity scores between comprehensive semantics of functions, the goal is to maximize the similarity of the functions with the same name (functions in the same blue dashed box) and minimize the similarity of functions with different names (functions in different blue dashed boxes).  
The utilization of this module is solely restricted to the training phase, where it facilitates the learning of comprehensive semantics of functions. It does not come into play during the testing process.

\subsection{Function Semantics Encoding}

To address \textbf{C1}, we construct comprehensive semantics encodings of functions by utilizing assembly instructions and control flow information. 
This module allows us to extract valuable semantic information from assembly functions. In this section, we explain how we generate the instruction representations using a pre-trained assembly language model. Additionally, we elaborate on the encoding of the fine-grained Control Flow Graph (CFG) using a K-hop Convolutional Graph Neural Network (ConvGNN) with multi-view node representation.

\noindent \textbf{Pre-trained Assembly Language Model.} \label{ALM}
To capture the underlying characteristics of assembly instructions, we develop a pre-trained assembly language model specifically designed for instruction representation learning. Our model is built upon BERT~\cite{b38}, and the input consists of a pair of instructions extracted from the control flow sequences or the data dependency (def-use) sequences. Each instruction is treated as a separate sentence, and each token within the instruction represents a word. 
We also apply normalization techniques to standardize the instructions.

Inspired by previous work~\cite{b40, b41, trex}, we employ three tasks to train our language model, aiming to capture the internal format, control flow dependencies, and data flow dependencies of instructions.
The first training task is Text Infilling, which is a specific noising task utilized in BART~\cite{b45}. In this task, we randomly select text spans and replace them with a single [MASK] token. The model is trained to predict the correct tokens that should appear in the masked positions. This task enhances the model's ability to understand the relationships between different parts of instructions.
The second training task is Control Dependency Inference (CDI). Since control flow does not have strict dependencies and order~\cite{b40}, the CDI task extends the control dependencies to instructions located $w$ steps before and after the target instruction within the same basic block. The model then predicts whether two given instructions co-occur within this range. This allows the model to learn the relationships between instructions, taking into account the specific characteristics of control flow.
The third training task is Def-Use Inference (DUI). Data flow dependencies tend to exhibit stability across different optimization levels. DUI predicts whether two given instructions have a data dependency, i.e., whether a later instruction uses a variable defined by the previous instruction. By incorporating def-use information into the instruction embeddings, this task enables the model to capture the relationships between instructions that rely on shared data. This comprehensive understanding of assembly function behavior facilitates effective representation learning.
\begin{figure}[tb]
    \vspace{-8pt}
    \centering
    \includegraphics[scale=0.28]{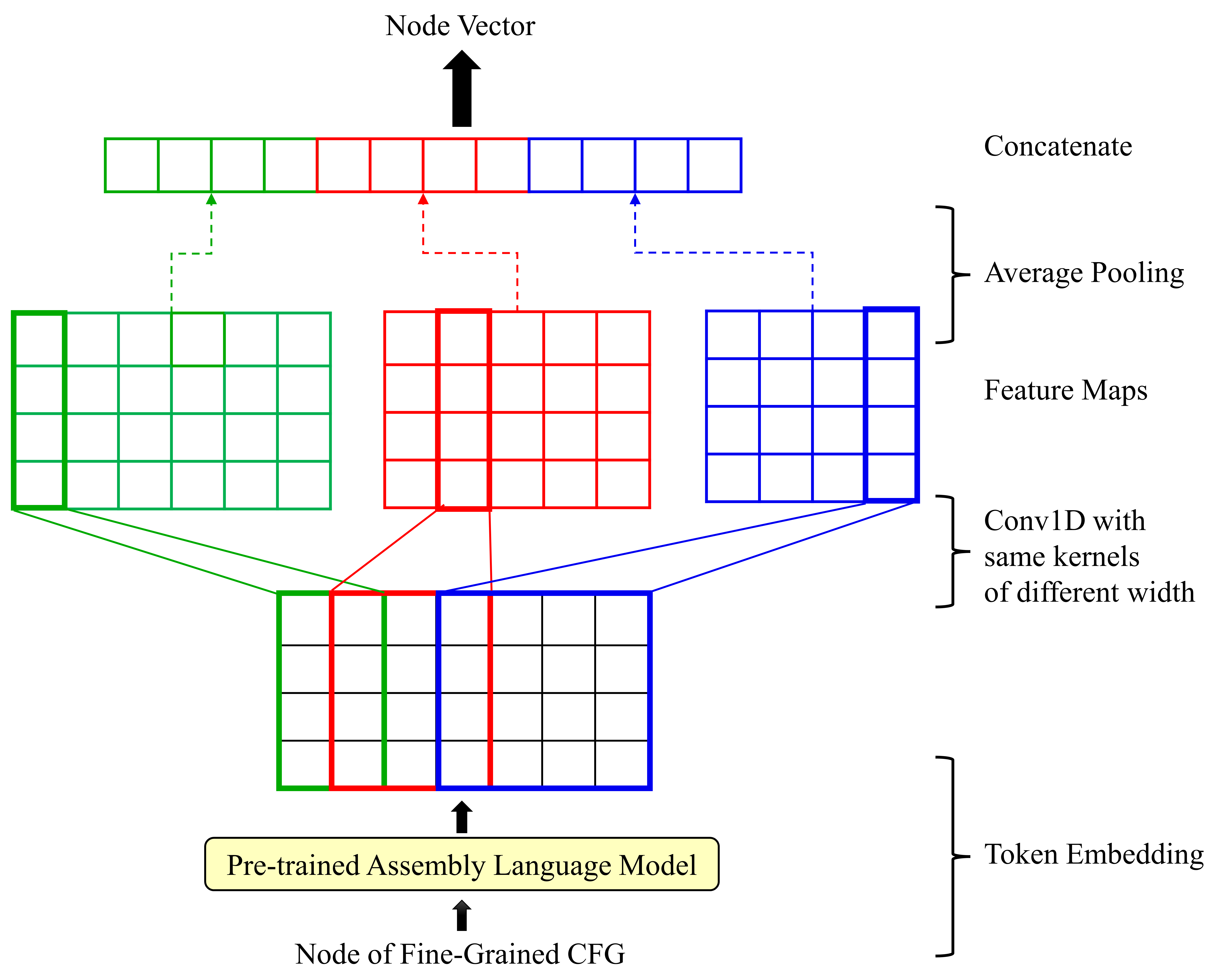}
    \caption{CNN for Generating a Node Vector. Using 4-Dimensional Embeddings, and the Widths of the Kernels Are 2, 3, and 4, Respectively. Feature Maps of Different Kernel Widths Are Marked in Different Colors.}
    \Description{ In This Example, We Use 4-Dimensional Embeddings, and the Widths of the Kernels Are 2, 3, and 4, Respectively. And concatenate the vectors. Feature Maps of Different Kernel Widths Are Marked in Different Colors}
    \label{fig:node_vec}
    \vspace{-12pt}
\end{figure}

\noindent \textbf{K-hop Convolutional Graph Neural Network.} \label{K-hop}
Previous research~\cite{b19,b20,b21} in source code summarization has demonstrated the effectiveness of neural networks in extracting information from graphs, as opposed to sequential token sequences. 
To enhance the modeling of assembly code, \textsc{Epitome} leverages the structural information provided by CFG.
However, a basic block in the CFG may contain a lot of tokens, which exceed the token limitations (e.g., 512 tokens) imposed by most existing models for sequence handling~\cite{b37}. 
To overcome this limitation and better capture the semantics of the CFG, we construct a fine-grained CFG. 
In this fine-grained CFG, nodes represent instructions, while edges depict jumps or sequential control flows.
To encode the fine-grained CFG, we propose a K-hop ConvGNN with a multi-view node representation. 
The network is introduced from two aspects of multi-view node representation and message passing framework.

\paragraph{Multi-View Node Representation.} \label{node_rep}
A straightforward approach to obtaining the representation of a node is to average or sum its token embeddings. However, this simplistic approach overlooks the importance of positional information within sequences. To address this limitation, we propose the utilization of one-dimensional convolution as a solution.
Figure~\ref{fig:node_vec} depicts our approach to generating a node vector. 
Let $ \mathcal{E}^i = (E_1, \ldots, E_m) \in \mathbb{R}^{d \times m} $ represent the embedded token sequence of node $i$, obtained using the pre-trained assembly language model. 
We perform an operation between $\mathcal{E}^i$ and convolution kernel $ \mathcal{K} \in \mathbb{R}^{d \times w} $ with the width $w$, to compute a feature map $ F^i \in \mathbb{R}^{l-w+1}$.
Specifically, the $j$-th element of $F^i$ is given by
\begin{equation}
     F_j^i = \mbox{ReLU}  \left( \sum {( \mathcal{K} \bigotimes \mathcal{E}_{j:j+w-1}^i) + b_{conv}} \right)
\end{equation}
where $ \mathcal{E}_{j:j+w-1}^i $ represents the $j$-th to $(j + w -1)$-th column of $\mathcal{E}^i$, $ b_{conv}$ denotes a bias vector, and the ReLU refers to activation function~\cite{nair2010rectified}. 
Following the convolution operation, we utilize average pooling to compute the feature of node $i$ associated with $ \mathcal{K}$:
\begin{equation}
    x_{ \mathcal{K}}^i = \frac{1}{l-w+1} \sum_{j=1}^{l-w+1} {F_j^i}
\end{equation}

To capture features of different n-grams, we utilize multiple kernels with varying sizes. These features are subsequently concatenated to form the node vector $x_i = (x^i_{{ \mathcal{K}}_1}, \ldots , x^i_{{\mathcal{K}}_n}) \in \mathbb{R} ^ n $, where $n$ is the total number of kernels.

\paragraph{Message Passing Framework.}
In the conventional message passing framework of ConvGNN, node representations are updated based on the direct neighbors of the node, also referred to as 1-hop neighbors~\cite{wu2020comprehensive}. 
However, the expressive power of 1-hop message passing has its limitations. 
The K-hop message passing has been introduced, which expands the scope of information exchange beyond direct neighbors.
The K-hop ConvGNN model consists of a series of neighborhood aggregation layers, where each layer generates updated representations for the nodes by aggregating information from all the neighbors within K hops of the nodes.

Denote a graph as $G = (V, E)$, where $V$ is the node set and $E \subseteq V \times V$ is the edge set. Each node $v \in V$ is associated with an initial feature vector $h^{0}_v \in \mathbb{R}^d$, which typically corresponds to the node representation $x_v$ mentioned above. We denote $\mathcal{N}_v^k$ as the K-hop neighborhood of a node $v$, which is the set of nodes at a distance less than or equal to $k$ from $v$. Suppose the ConvGNN model with $L$ neighborhood aggregation layers. At the $l$-th neighborhood aggregation layer ($l > 0$), the K-hop message passing framework as follow:
\begin{align}
    M^{l,k}_v &= \mbox{MESSAGE}^{l}_k \left( \left\{ h^{l-1}_u |\  u \in \mathcal{N}_v^{k}\right\}\right)\\
    h^{l,k}_v &= \mbox{UPDATE}^{l}_k \left( M^{l, k}_v , \ h^{l-1}_v \right) \\
    h^{l}_v &= \mbox{COMBINE}^{l} \left( \left\{ h^{l,k}_v \ |\ k=1,2,\ldots K \right\}\right)
\end{align}
where $M^{l,k}_v$ represents the aggregated message from the $k$-th hop neighbors to the node $v$ at layer $l$, $h^{l,k}_v$ as the $k$-th hop output representation of node $v$ at layer $l$.
$\mbox{MESSAGE}^{l}_k$ and $\mbox{UPDATE}^{l}_k$ are message and update functions at layer $l$ respectively. 
The \mbox{COMBINE} function is utilized to combine the representations of node $v$ at different hops. The output representation of node $v$ at layer $l$ is denoted as $h^{l}_v$. After $L$ layers of message passing, $h^{L}_v$ is employed as the final representation of node $v$. In order to obtain the graph representation $h_G$, a READOUT function is applied.
\begin{equation}
    h_G = \mbox{READOUT} \left\{ h^{L}_v \ |\ v \in V \right\}
\end{equation}

To address the issue of excessively similar node representations in ConvGNN as the number of layers increases, we enhance the local representativeness of individual nodes. 
We achieve this by encoding normalized instructions using a standard Transformer encoder, resulting in a vector representation denoted as $h_{inst}$. 
These encoded instructions are then concatenated with the graph representation $h_{G}$ to form the semantic encoding of the function $emb = [h_{G}; h_{inst}]$.

\subsection{Function Name Prediction Module} \label{MTLM}

As shown in Figure~\ref{fig:arch}, our function name prediction model follows the encoder-decoder paradigm. The semantic encoding of the function serves as the input to the decoder. 
Our model aims to predict $P(Y|X)$, where $X$ represents the semantic encoding of the function and $Y = \{y_1, \ldots, y_m\}$ represents the corresponding function name labels. The decoder, which consists of a standard transformer decoder, generates the output $O_t$.
The generation probabilities of $t$-th label can be calculated by
\begin{equation}
    P_t = P(y_t|\hat{Y}_{1:t-1}; X) = softmax(U^t \cdot O_t + b^t) 
\end{equation}
where $U^t$ denotes projection parameters, $b^t$ is a bias term. $\hat{Y}_{1:t-1}$ denotes the labels generated up to time $t-1$. 
$P_t$ represents the probability distribution of the generated word at time $t$.

To optimize the generation probability distributions of the function name, we minimize the following maximum likelihood loss functions:
\begin{equation}
    \mathcal{J}_{cg}(\theta_{cg}) = - \sum_{t=1}^{m}{\log ( P_t)}
\end{equation}
where $m$ is the size of the generated labels and $\theta_{cg}$ is the parameters set for this task.

\subsection{Function Semantics Similarity Prediction Module}
In our approach, the function semantics similarity prediction task is used as an auxiliary task to tackle \textbf{C2}.
By sharing the encoder between the two tasks, the function semantics encoding module can learn semantic knowledge that applies to different optimization levels binaries.
Taking inspiration from the widely used Relief algorithm~\cite{kononenko1994estimating, robnik2003theoretical} in feature selection~\cite{guyon2003introduction, james2013introduction, Liu2010}, this task learns function semantics based on the similarity between functions belonging to the same or different names.
The similarity score of two function $(\mathcal{F}_1, \mathcal{F}_2)$ is defined as:
\begin{align}
    Score(\mathcal{F}_1, \mathcal{F}_2) &=f(h_1, h_2) \\
     f( h_1, h_2) &= \mbox{cos} ( W_{h_1} h_1 + b_{h_1}, W_{h_2} h_2+ b_{h_2}) \\
     h_1 &= \mbox{tanh} (maxpooling(emb_1)) \\
    h_2 &= \mbox{tanh} (maxpooling(emb_2))
\end{align}
where $f$ is the cosine similarity function, $h_1$, $h_2$ is calculated by the function semantics encoding module, $W_{h_1}$,$b_{h_1}$, $W_{h_2}$, $b_{h_2}$ are trainable parameters.

To define the training objective, we utilize the Ranking Loss~\cite{schroff2015facenet}. The goal of our training is to maximize the similarity between the paired sample ($\mathcal{F}_x$, $\mathcal{F}_y$) originating from the same source code but with different optimization levels, while minimizing the similarity between the pair ($\mathcal{F}_x$, $\mathcal{F}_z$), where $\mathcal{F}_z$ is randomly chosen and has a different name from $\mathcal{F}_x$. We aim to ensure that the difference between the similarities of ($\mathcal{F}_x$, $\mathcal{F}_y$) and ($\mathcal{F}_x$, $\mathcal{F}_z$) is at least greater than a margin $M_{cs}$. Therefore, the loss function can be defined as:
\begin{equation}
    \mathcal{J}_{cs}(\theta_{cs}) = max( f( h_x, h_y) - f( h_x, h_z)- M_{cs}, 0) 
\end{equation}
where $\theta_{cs}$ represents the collective parameters for the function semantics similarity prediction task. 

To enhance the shared encoder, we train the two related tasks jointly. Specifically, we optimize the two tasks by minimizing the following objective function:
\begin{equation}
    \mathcal{J}(\Theta) = \hat{\lambda}_1 \mathcal{J}_{cg} + \hat{\lambda}_2 \mathcal{J}_{cs}
\end{equation}
where $\Theta$ represents the set of parameters of our model, $\hat{\lambda}_1$ and $\hat{\lambda}_2$ represent the hyperparameters controlling the importance of the corresponding objective functions.

\section{Dataset Construction}
In this section, we provide a detailed explanation of the dataset construction method used to alleviate \textbf{C3}, which involves addressing the label sparsity issue and the OOV problem through two data preprocessing approaches.

\subsection{Dataset Overview} \label{dataset}
Our dataset consists of 174 open-source projects obtained using the multi-platform cross-compilation tool, \textit{Buildroot}~\cite{buildroot}. These projects include widely used projects in previous research~\cite{b44, b7, yang2021asteria}, such as coreutils~\cite{coreutils} and binutils~\cite{binutilss}. We compile these projects into four architectures (x86, x64, ARM, and MIPS) and five optimization levels (O0, O1, O2, O3, and Os). The final dataset comprises 36,275 binaries and 2,597,346 functions from these binaries, as shown in Table~\ref{tab:firm}.

To ensure a reliable evaluation of \textsc{Epitome}, the five-fold cross-validation technique is employed. Following the precedent set by analogous experiments in function name prediction research~\cite{b4, b5, b44}, our methodology involved partitioning the dataset into train-validation-test sets using an 8:1:1 ratio for each individual run. 
To address the adverse impact of code duplication in open-source datasets~\cite{b4, b8, b9}, we categorized datasets into training, validation, and test sets based on source code. Binaries from the same source code, compiled with different optimization levels, were then assigned to one of the sets. This ensures that binaries from identical source code do not concurrently appear in the sets.

\begin{table}[tb]
    \centering
    \vspace{-6pt}
    \caption{Number of Projects, Binaries and Functions in Our Datasets}
    \begin{center}
    \begin{tabular}{cccc}
    \toprule
         {\textbf{Architecture}}&{\textbf{\# of projects}}&{\textbf{\# of binaries}}&{\textbf{\# of functions}} \\
         \midrule
         {x86} &{174}&{9,110}&{646,759}\\
         
         {x64}&{174}&{9,075}&{629,403}\\
         
         {ARM}&{174}&{9,075}&{673,246}\\
         
         {MIPS}&{174}&{9,015}&{647,938}\\
         \hline
         {total}&{696}&{36,275}&{2,597,346}\\
    \bottomrule
    \end{tabular}
    \vspace{-10pt}
    \label{tab:firm}
    \end{center}
\end{table}

\subsection{Votes-Based Function Name Tokenization} \label{tokenize}
Function names that adhere to standard naming conventions can be tokenized using uppercase letters, underscores, or numbers, as shown in Table~\ref{tab:name_method}. 
However, existing tokenization tools~\cite{Wordninja, seeha-etal-2020-thailmcut} often struggle to accurately segment uncommon names, particularly when domain-specific jargon is involved.
To tackle this challenge, we propose the votes-based function name tokenization method. This approach combines an unsupervised learning tokenization model, a unigram language model, and a rule-based algorithm to tokenize function names into meaningful labels. To build the function name corpus, we utilize data from CodeSearchNet~\cite{husain2019codesearchnet} and an API names dataset~\cite{demirkiran2022ensemble}.

\noindent \textbf{Unsupervised Learning Tokenization Model.}  
The model is based on N-gram-to-symbol counts, which analyze the frequency of single symbols that follow or precede every possible N-gram. The frequency is referred to as the transition freedom (TF). The model comprises three components:
\begin{itemize}
    \item Counts of N-grams encountered throughout the corpus, which form the lexicon dictionary.
    \item Counts of all symbols appearing after each specific N-gram, representing the forward transition freedom.
    \item Counts of all symbols appearing before each specific N-gram, representing the backward transition freedom.
\end{itemize}

Label boundaries are set using peak TF values, calculated as the difference between a transition's TF and the preceding one's~\cite{kolonin2022unsupervised}, and are doubled if the position matches a lexicon entry.

\noindent \textbf{Unigram Language Model.} 
The model assumes that the probability of a label is independent of its surrounding labels. The probability of a function name can be calculated by multiplying the probabilities of its individual labels, where the probabilities of labels are determined by their frequencies in the corpus. We employ the forward maximum match strategy~\cite{yan2021hmm} to tokenize the function name.

\noindent \textbf{Rule-Based Algorithm.} 
We begin by constructing a predefined list of frequently used programming abbreviations and specialized domain-specific jargon, such as \textrm{msg} for message and \textrm{lst} for list. Next, we utilize a dynamic programming algorithm to tokenize function names into non-overlapping labels. 
Our goal is to identify the longest combination of English words or manually defined program words among all possible label permutations. For instance, when tokenizing the name "timeset," our algorithm would produce \{time, set\} instead of \{times\}.

By combining the strengths of the rule-based algorithm, unigram language model, and unsupervised learning tokenization model, we can effectively tokenize function names into meaningful labels. The rule-based algorithm is particularly effective in tokenizing short words, while the unigram language model and unsupervised learning tokenization model excel in preserving long words. The unigram language model focuses on single words, whereas the unsupervised learning tokenization model considers the contextual relationship between labels. 
To determine the final tokenization position list, we combine the three tokenization position lists and apply the rules:
\begin{itemize}
    \item If at least two of the three models suggest adding a position to the final list, it is included.
    \item The pending list is determined by selecting the list with the largest intersection with the final list.
    \item The final label boundaries are obtained by taking the union of the pending list and the final list, with any overlapping parts of the labels removed.
\end{itemize}

This approach effectively addresses the issues of label sparsity and OOV by tokenizing function names into meaningful labels. Furthermore, we enhance the informativeness of the labels by expanding abbreviations using a predefined list of common programming abbreviations.

\subsection{Label Relationships Identification} 
Different developers may use different words to name the same function, which can contribute to label sparsity. Ignoring the close relationship between various forms of the same word can exacerbate this issue.
In situations where function names contain abbreviations or domain-specific jargon, identifying relationships among labels becomes challenging. 
To address this, we draw inspiration from prior works~\cite{b6, chen2019sethesaurus, b44} and adopt a set of approaches to identify semantically related labels and establish associations among them.

To establish semantic relationships among labels, we employ various techniques depending on the nature of the labels. For English words, we utilize stemming and lemmatization techniques to transform different forms of a word into its original form.
For other labels, we leverage two widely used word embedding models: Skip-Gram~\cite{mikolov2013efficient} and FastText~\cite{bojanowski2017enriching}. The Skip-Gram model learns vector representations of context words given a central word within a fixed-size window. FastText model considers n-gram sets of a word as input, allowing it to generate vector representations even for rare words, terms, and OOV words.
Using these models, we obtain vector representations for each label in the function name vocabulary. We then select the top-$10$ labels with the highest similarity in the vector space $V$ based on the Skip-Gram and FastText models. These 20 labels form the set of candidate semantically related labels.

In the list of candidate semantically related labels, we categorize them into three types: synonyms, abbreviations, and other related words. For synonyms, we utilize the Smith-Waterman algorithm~\cite{smith1981identification} to calculate the similarity between the query label $t$ and each candidate label. We set a threshold of relative similarity at $\frac{2}{3}$ to determine if a candidate label is a synonym.
If a candidate label is not a synonym, we proceed to check if it is an abbreviation for the given label $t$. We consider two labels to have an abbreviation-full name relationship if $t$ starts with $w$ or $w$ starts with $t$.
Finally, we use WordNet~\cite{wordnet} to identify labels that have other relationships with the label $t$ from the candidate set.
To ensure the quality of the selected labels, we conduct manual verification, similar to the approach employed in existing work~\cite{chen2019sethesaurus, b6, b44}. This step helps eliminate noisy labels from the candidate set, ensuring the reliability of the identified semantically related labels. By successfully establishing associations between labels, we effectively alleviate the label sparsity issue.

\section{EVALUATION}
In this section, we conduct extensive experiments to answer the following research questions:
\begin{itemize}
    \item \textbf{RQ1:} How effective is \textsc{Epitome} in function name prediction?
    \item \textbf{RQ2:} How does \textsc{Epitome} compare to the state of the art?
    \item \textbf{RQ3:} How does each component of \textsc{Epitome} contribute to its efficacy?
    \item \textbf{RQ4:} Do the issues of label sparsity and OOV indeed affect the performance of \textsc{Epitome}?
\end{itemize}

\subsection{Experimental Setup}

\noindent \textbf{Environment.}
We conducted our experiments on a machine equipped with a 3.0 GHz Intel Xeon Gold 6248R CPU, 1024 GB RAM, and 4*NVIDIA RTX-3090 GPUs. The software platforms used include Ubuntu 18.04 and Python 3.8.
For the extraction of functions from binaries, we utilized IDA Pro 7.3~\cite{b3} along with the LLVM IR plugin~\cite{b43}. 
To preprocess the function names, we employed NLTK~\cite{bird2009natural} and python-Levenshtein~\cite{b27} libraries in Python.
The implementation of \textsc{Epitome} was done using Python language with the support of PyTorch~\cite{paszke2019pytorch} and Gensim~\cite{rehurek2011gensim}.

\noindent \textbf{HyperParameters.}
The dimension of token embeddings is 128, the encoder consists of the 6-layer and 8-head attention layer with 256 hidden states, and the ConvGCN consists of the 2-layer encoder with 256 hidden states. The dropout is 0.1, and the batch size is 32. 
Then we trained the model by Adam optimizer~\cite{b28} with an initial learning rate of $5 \times 10^{-5}$. 

\noindent \textbf{Evaluation Metrics.}
In the evaluation of \textsc{Epitome} and existing techniques, we adopted classical precision, recall, and F1-score metrics at the word-level, as inspired by related studies~\cite{b4, b6, b7, b29}.
To calculate these metrics, we define the following variables.
True Positives (\textit{TP}), is the number of labels correctly predicted in the function name. 
False Positives (\textit{FP}), is the number of incorrectly predicted labels in the function name. 
False Negatives (\textit{FN}), is the number of labels that actually exist but were not predicted by \textsc{Epitome}. 
More precisely, given a predicted function name
$ \hat{Y}: \{\hat{y_1}, \ldots, \hat{y_j}\}$ and the ground truth $Y: \{y_1, \ldots, y_k\}$, where $y_i (\hat{y_i})$ is a label in function name, $\alpha$ is the indicator function, $\alpha \{True\} =1$ and $\alpha \{False\} = 0$.
\begin{equation}
    TP = \sum_{i=1}^j \alpha {(\hat{y_i} \in Y)} \quad
    FP = \sum_{i=1}^j \alpha {(\hat{y_i} \not\in Y)} \quad
    FN = \sum_{i=1}^k \alpha {({y_i} \not\in \hat{Y})}
    \label{TP}
\end{equation}
and we compute precision (P), recall (R), and F1-score (F1) as:
\begin{align}
    P = \frac{TP}{TP+FP} \qquad 
    R = \frac{TP}{TP+FN} \qquad
    F1 = \frac{2 \times P \times R}{P + R}
\end{align}

\subsection{Overall Effectiveness}

Our analysis began with a thorough evaluation of \textsc{Epitome}'s performance using datasets prepared through a five-fold cross-validation method. The comprehensive performance metrics for \textsc{Epitome} are summarized in Table~\ref{tab:overall}. \textsc{Epitome} exhibits varying levels of performance across different architectures, particularly in terms of precision, recall, and the F1 score.
\textsc{Epitome} demonstrates commendable performance in the weighted macro category, achieving a precision of 73.13\%, a recall of 70.84\%, and an F1 score of 71.96\%.
It is worth noting that, when subject to different compilation optimization levels within the same architecture, the F1 score of \textsc{Epitome} maintains a consistent level of stability. However, an exception to this pattern is observed in the case of the ARM dataset.

We observe that the performance of \textsc{Epitome} is better on O1-Os binaries compared to O0 binaries. 
Upon manual inspection, we discover that O0 binaries contain a higher number of functions compared to O1-Os binaries, despite being derived from the same projects. This discrepancy in function count may explain the lower performance of O0 optimization. The presence of numerous small and similar functions at the O0 optimization level can confuse our model and lead to inaccurate function name predictions.
However, by utilizing function inlining, which is enabled for O1-Os optimizations as specified in the GCC manual~\cite{gcc}, the number of these functions decreases in O1-Os binaries. This reduction helps mitigate the interference caused by small and similar functions, thereby improving the overall performance.
For instance, in the ARM dataset, when we remove functions from the O0 dataset to align it with the O3 datasets, the F1 scores of \textsc{Epitome} at O0 and O3 are 76.38\% and 76.61\%, respectively. 
This finding highlights the substantial negative impact of having a greater number of small and similar functions on the performance of \textsc{Epitome}.

\begin{table}[tb]
    \centering
    \vspace{-8pt}
    \caption{Overall Performance Across Different Architectures (Arch) and Optimizations (Opt)}
    \begin{center}
    \resizebox{\columnwidth}{!}{
    \begin{tabular}{ccccc|ccccc}
    \toprule
         {\textbf{Arch}}&{\textbf{Opt}}&{\textbf{Precision}}&{\textbf{Recall}}&{\textbf{F1 Score}} &{\textbf{Arch}}&{\textbf{Opt}}&{\textbf{Precision}}&{\textbf{Recall}}&{\textbf{F1 Score}} \\
         \midrule
         \multirow{5}{*}{x86} &{O0}&0.7206&0.7024&0.7114& \multirow{5}{*}{x64}&{O0}&0.7021&0.6807&0.6912\\
         &{O1}&0.7534&0.7267&0.7397& &{O1}&0.7301&0.7050&0.7172\\
         &{O2}&0.7564&0.7281&0.7419&  &{O2}&0.7339&0.7094&0.7214\\
         &{O3}&0.7637&0.7349&0.74894& &{O3}&0.7393&0.7153&0.7274\\
         &{Os}&0.7488&0.7235&0.7359& &{Os}&0.7338&0.7093&0.7213\\
         \hline
         \multirow{5}{*}{ARM}&{O0}&{0.6807}&{0.6691}&0.6748& \multirow{5}{*}{MIPS}&{O0}&0.7018&0.6966&0.6992\\
         &{O1}&{0.7356}&{0.7106}&0.7228& &{O1}&0.7154&0.0.6914&0.7030\\
         &{O2}&{0.7384}&{0.7144}&0.7261& &{O2}&0.7260&0.7030&0.7142\\
         &{O3}&{0.7464}&{0.7205}&0.7331& &{O3}&0.7345&0.7096&0.7217\\
         &{Os}&{0.7379}&{0.7148}&0.7260& &{Os}&0.7275&0.7046&0.7158\\
    \bottomrule
    \end{tabular}
    }
    \label{tab:overall}
    \vspace{-12pt}
    \end{center}
\end{table}

\begin{tcolorbox}
\noindent \textbf{RQ1 Answer:} \textsc{Epitome} achieves 73.13\% precision, 70.84\% recall, and 71.96\% F1 score across different architectures and optimization levels, which is effective in predicting binary function names. 
\end{tcolorbox}

\subsection{Baseline Comparison}
We conducted a comparison of \textsc{Epitome} with two state-of-the-art binary function name prediction tools, NFRE~\cite{b6} and SymLM~\cite{b44}. Both NFRE and SymLM demonstrated superior performance compared to Debin~\cite{b5} and Nero~\cite{b4}, according to their reported results. However, we encountered the same issue as SymLM when evaluating Punstrip~\cite{b7}, it is currently unavailable. Therefore, we have excluded the evaluation results of Debin, Nero, and Punstrip from our comparison. 

\begin{figure}[bt]
    \centering
    \begin{minipage}{0.5\linewidth}
		\centering
		\includegraphics[width=0.95\linewidth]{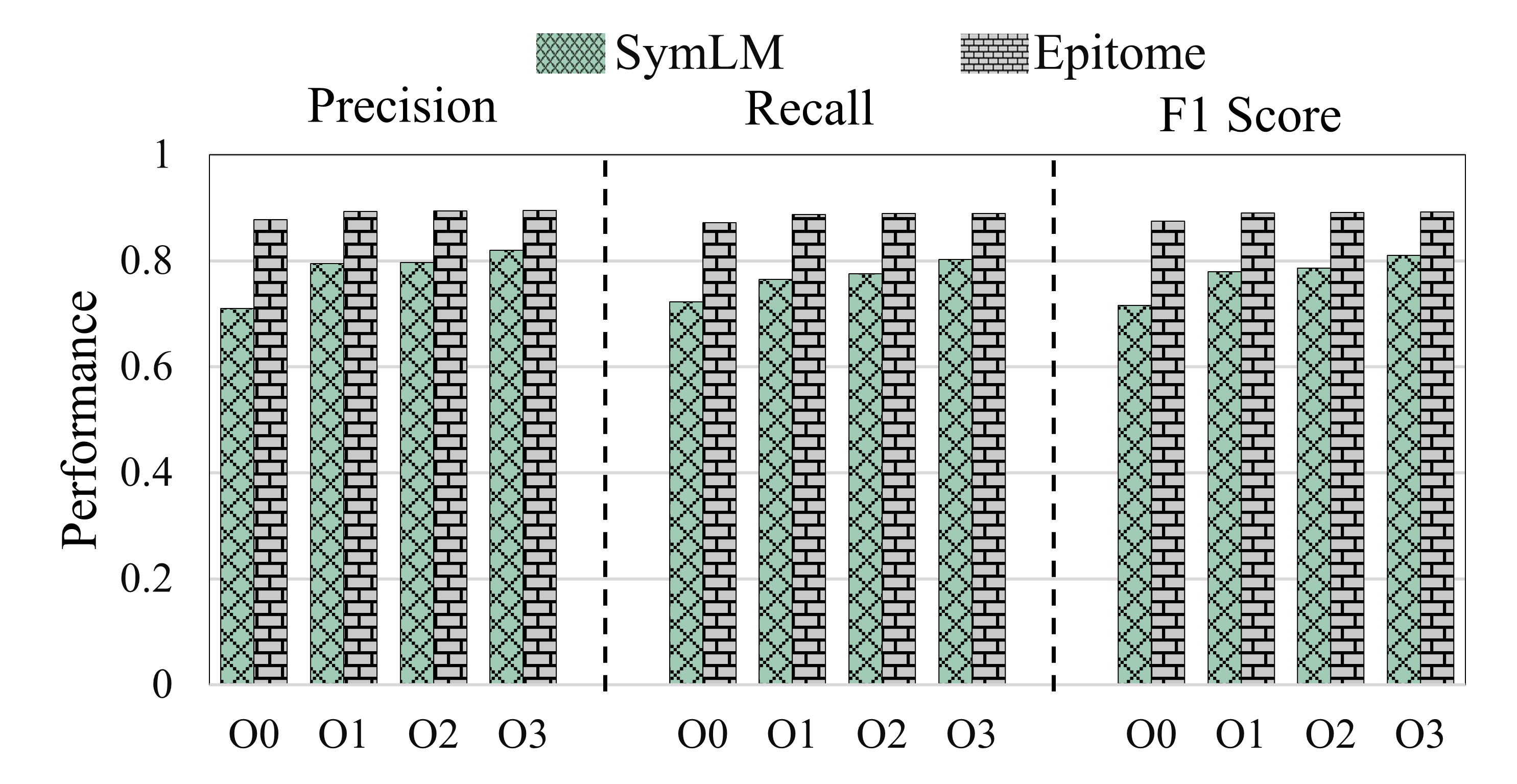}
		\caption{Comparison With SymLM on SymLM's Dataset.}
		\label{fig:compv1}
	\end{minipage}
        \begin{minipage}{0.49\linewidth}
		\centering
		\includegraphics[width=0.95\linewidth]{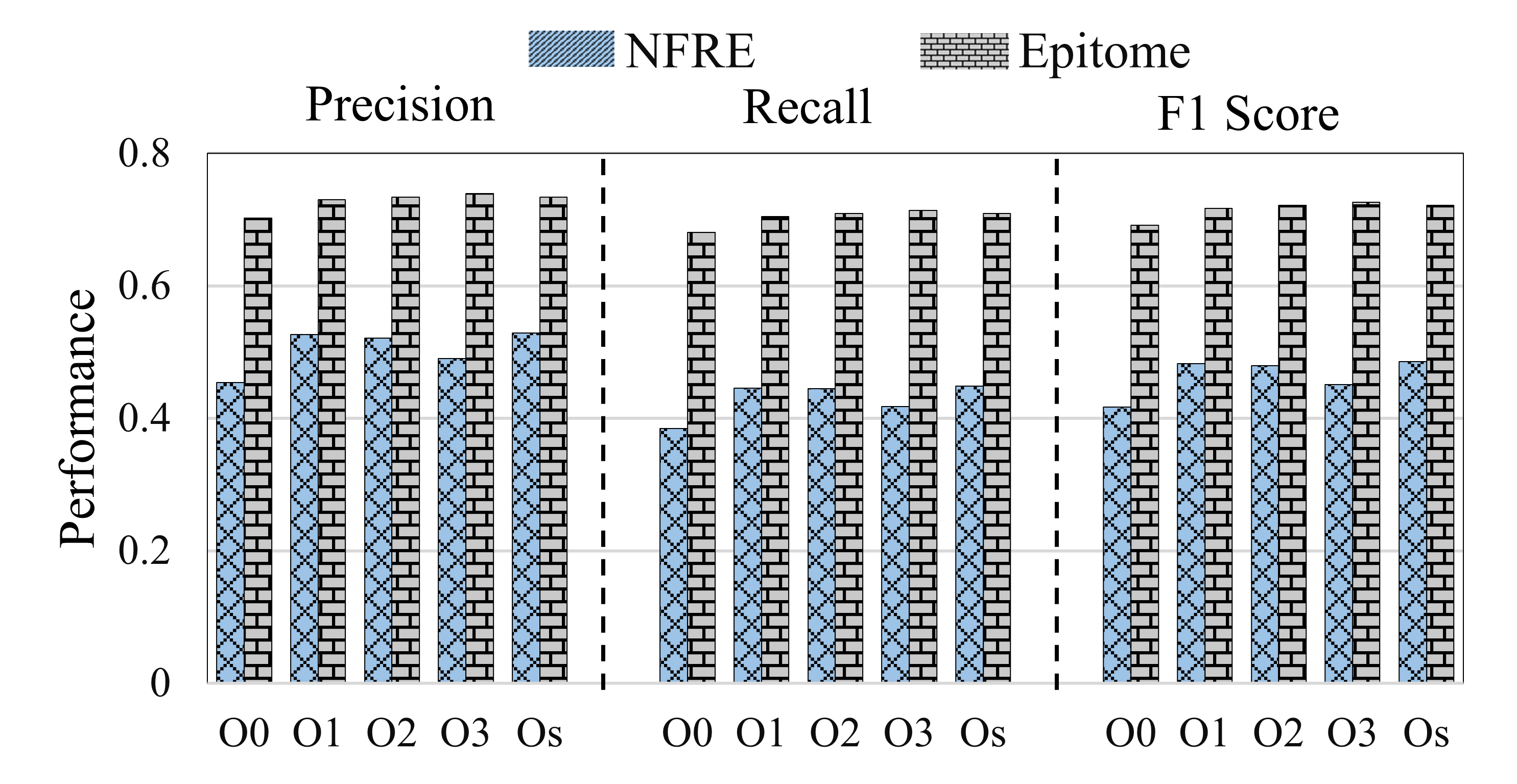}
		\caption{Comparison With NFRE on Our x64 Dataset.}
		\label{fig:compv2}
	\end{minipage}
    \vspace{-12pt}
\end{figure}

\noindent \textbf{Comparison with SymLM.}
We accessed SymLM's code on GitHub~\footnote{\url{https://github.com/OSUSecLab/SymLM}} and reached out to the authors to obtain the dataset containing x64 binaries compiled with four different optimization levels (O0-O3).
We employed a five-fold cross-validation approach to train and evaluate both SymLM and \textsc{Epitome} on SymLM's dataset, adhering to hyperparameter settings for SymLM as detailed in its associated publications. Both models were trained for the same number of times and adopted an early stopping strategy.
Figure~\ref{fig:compv1} illustrates the evaluation results, demonstrating that \textsc{Epitome} outperforms SymLM by 14.10\%, 15.40\%, and 14.75\% in precision, recall, and F1 score, respectively. Specifically, SymLM achieves an average performance of 0.780 precision, 0.766 recall, and 0.773 F1 score, while \textsc{Epitome} achieves an average performance of 0.890 precision, 0.884 recall, and 0.887 F1 score.
These results indicate that SymLM's performance varies significantly depending on the optimization level used, whereas \textsc{Epitome} maintains stable performance across different optimization levels. This highlights the strong performance of \textsc{Epitome} even when confronted with binaries of unknown optimization levels.

\noindent \textbf{Comparison with NFRE.}
NFRE does not provide code in its GitHub~\footnote{\url{https://github.com/USTC-TTCN/NFRE}}, so we reached out to the authors for further details. To evaluate NFRE, we followed the preprocessing steps recommended by the authors and applied them to our x64 binaries.
We utilized a five-fold cross-validation approach to train and evaluate both NFRE and \textsc{Epitome}. 
Most of the hyperparameters adhere to those specified in its publication, yet we conducted a grid search to refine certain parameters—such as employing dropout of 0.1, batch size of 128, and learning rate of 0.01—aiming to optimize its performance specifically for our dataset. This meticulous tuning was carried out to ensure a fair comparison with the evaluation outcomes presented by \textsc{Epitome}.
Illustrated in Figure~\ref{fig:compv2}, the evaluation results reveal \textsc{Epitome}'s superior performance over NFRE, showing enhancements of 44.34\% in precision, 64.16\% in recall, and 54.44\% in F1 score, respectively. 
One observation during the training of NFRE is that it converts instructions into standard representations, such as INST\_26d2 and INST\_26de, which lose the information of the instructions. Additionally, NFRE encodes the information of the CFG, which varies significantly under different optimization levels. This variability makes it challenging for NFRE to learn effective patterns for function name prediction.

\noindent \textbf{Generalizability Test.} ~\label{generalizability}
To evaluate \textsc{Epitome}'s generalizability, we executed two experimental sets. Our dataset encompassed diverse domains—cryptography, image processing, databases, compression, and USB operations—prompting us to assemble two distinct evaluation datasets. The first, a domain-knowledge shared set, comprised projects akin in theme but outside our dataset. Conversely, the second, named domain-knowledge deviated, intentionally included projects dissimilar to our dataset's specialized domains. We applied the previously trained \textsc{Epitome}, SymLM, and NFRE models to these datasets to gauge their adaptability across various domains.

We initiated our study by comparing the performance of NFRE and \textsc{Epitome}, both trained on our x64 dataset, on two evaluation sets. We achieved partially correct predictions for function names, detailed in Table~\ref{tab:proj}, highlighting \textsc{Epitome}'s superior performance over NFRE across both evaluation datasets. In the domain-knowledge shared evaluation dataset, \textsc{Epitome} significantly outperformed NFRE, achieving an accuracy improvement of 199.69\%.
This disparity suggests NFRE's potential difficulty in capturing complex semantics of functions, likely due to the loss of information from its instruction normalization process. 
In the domain-knowledge deviated evaluation dataset, both models experienced a notable decrease in performance. Nevertheless, \textsc{Epitome} maintained a significant lead, boasting a 61.19\% higher accuracy than NFRE. This underscores \textsc{Epitome}'s superior adaptability and generalization capacity when dealing with unfamiliar domains, compared to NFRE.

\begin{table}[bt]
    \centering
    \vspace{-6pt}
    \caption{Generalizability Testing of SymLM, NFRE and \textsc{Epitome} on Unknown Binary Functions.}
    \begin{center}
    \begin{tabular}{p{2.6cm}ccc|cc}
    \toprule
         \multirow{2}{*}{\textbf{Project Category}} & \multirow{2}{*}{\textbf{Domain Category}} &\multicolumn{4}{c}{\textbf{Accuracy}} \\
         \cline{3-6}
         &&\textbf{NFRE}&\textbf{\textsc{Epitome}} &\textbf{SymLM} & \textbf{\textsc{Epitome}} \\
         \midrule
         \multirow{5}{*}{\makecell{Domain-Knowledge \\ Shared}} &{Cryptography}&0.0638&0.4545&0.2348&0.4957 \\ 
         &{Image Processing}&0.2044&0.5258&0.2778&{0.4815} \\ 
         &{Databases}&0.1667&0.4528&0.5161&{0.5256} \\   
         &{USB Operations}&0.1682&0.4286&0.3125&{0.5408} \\ 
        &{Compression Algorithms}&0.1779&0.4789&0.3981&{0.4886} \\ 
         \hline
         \multirow{2}{*}{{\makecell{Domain-Knowledge \\ Deviated}}}&\multirow{2}{*}{Another Domain}& \multirow{2}{*}{0.0773}&\multirow{2}{*}{0.1246}&\multirow{2}{*}{0.1936}&\multirow{2}{*}{0.2258} \\
         &&&&& \\
    \bottomrule
    \end{tabular}
    \label{tab:proj}
    \end{center}
    \vspace{-12pt}
\end{table}

In assessing generalizability, we pitted \textsc{Epitome} against SymLM, both trained on SymLM's x64 dataset.
\textsc{Epitome} excelled in both evaluation datasets, markedly surpassing SymLM with a 45.59\% accuracy boost in domain-knowledge shared dataset and maintaining a 16.63\% lead in domain-knowledge deviated dataset, where performance decline was evident for both models.
We attribute this drop in accuracy to the substantial distribution shifts between the training dataset and the domain-knowledge deviated dataset.
Notably, the KL divergence~\cite{trex} between this evaluation dataset and the training dataset was $1.4\times$ greater than the divergence between the training dataset and domain-knowledge shared dataset.
This substantial shift in distribution likely contributed to the reduced performance of both models. 
The results highlight \textsc{Epitome}'s exceptional ability to generalize across previously unseen binaries, illustrating its adaptability and robustness in navigating unknown binaries. 
Notably, \textsc{Epitome} demonstrates enhanced performance with binaries that share domain-specific knowledge with those in its training dataset.

Additionally, a recent Synopsys~\cite{Synopsys} report reveals that an overwhelming 96\% of audited software applications now include at least one third-party library.
Given this trend, broadening our dataset to encompass a wider array of third-party libraries to enhance \textsc{Epitome}'s effectiveness in predicting names for unknown binary functions.

\begin{tcolorbox}
    \noindent \textbf{RQ2 Answer:} \textsc{Epitome} is more effective than state-of-the-art works. For example, it outperforms SymLM by 14.10\%, 15.40\%, and 14.75\% on precision, recall and F1 score. \textsc{Epitome} has better generalizability than state-of-the-art works. For example, it surpasses NFRE in performance by 199.69\% on the evaluation dataset that shares domain knowledge, and it exceeds NFRE by 61.19\% in accuracy on the domain-knowledge deviated evaluation dataset.
\end{tcolorbox}

\subsection{Ablation Study}
\begin{figure}[bt]
    \centering
    \vspace{-6pt}
    \begin{minipage}{0.45\linewidth}
		\centering
		\includegraphics[width=0.85\linewidth]{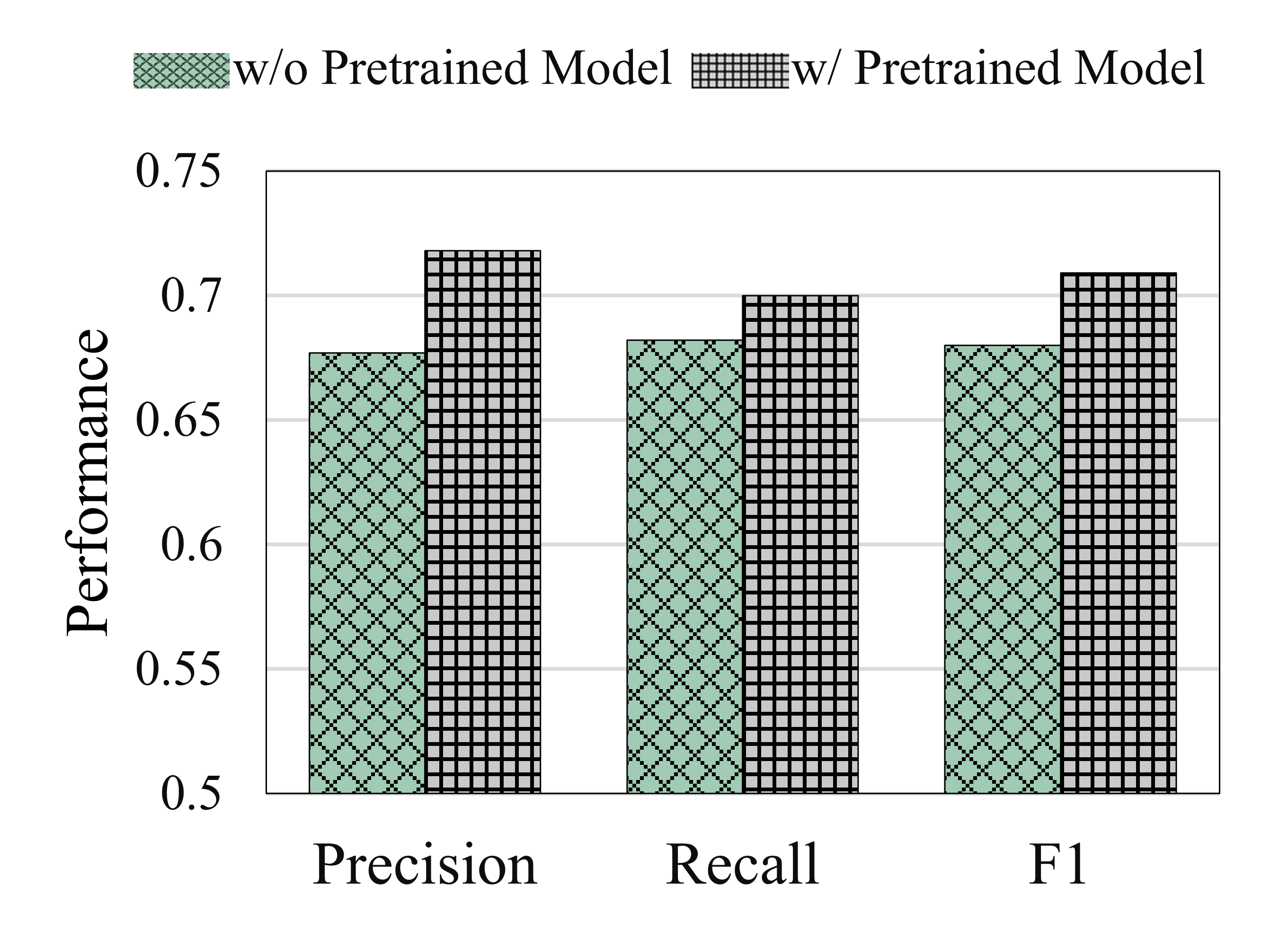}
		\caption{Effectiveness of Pre-trained Model.}
		\label{fig:nopre}
	\end{minipage}
     \vspace{-5pt}
        \begin{minipage}{0.45\linewidth}
		\centering
		\includegraphics[width=0.85\linewidth]{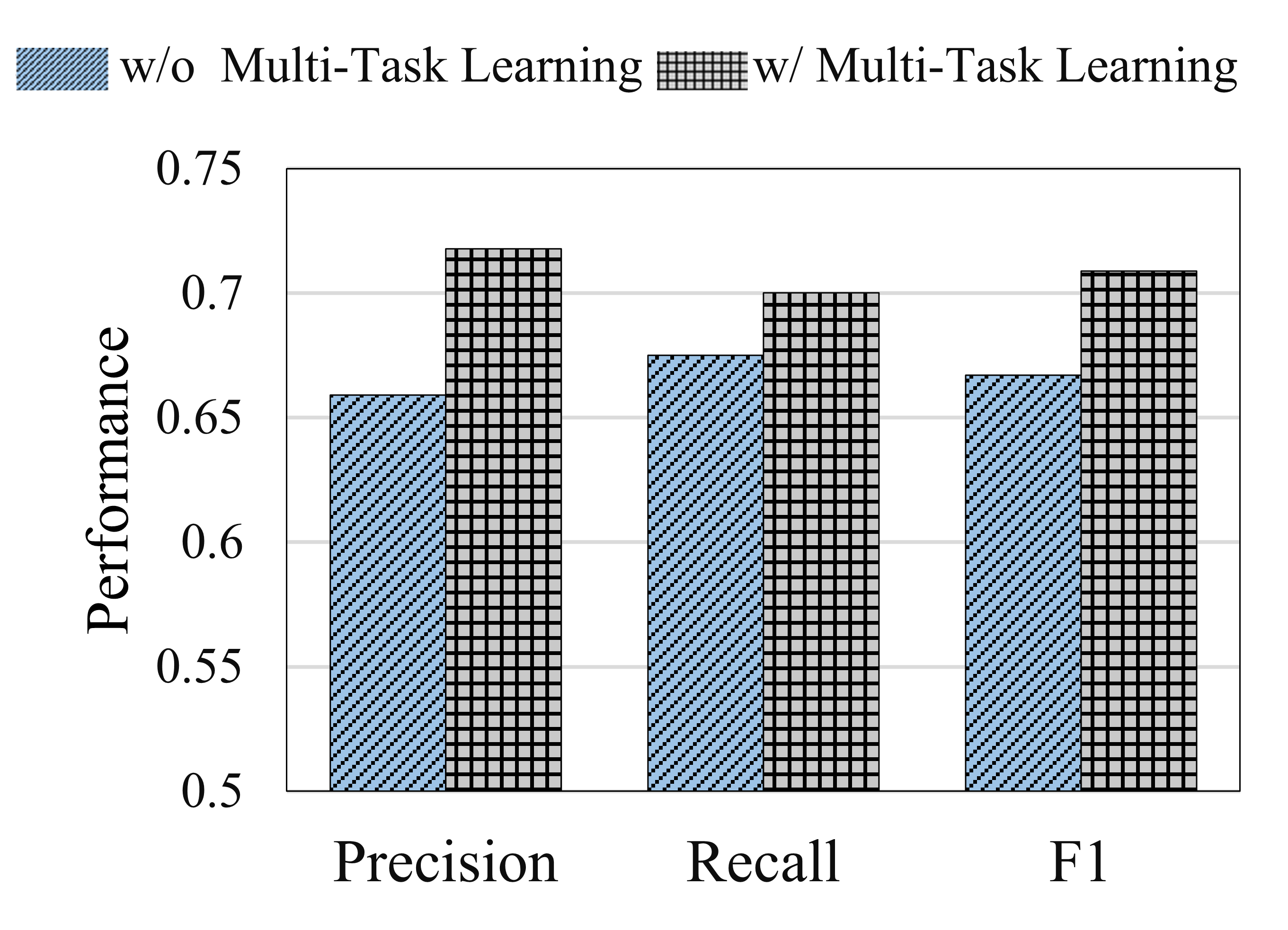}
		\caption{Effectiveness of Multi-Task Learning.}
		\label{fig:mul-task}
	\end{minipage}
    \vspace{-5pt}
\end{figure}

In this section, we endeavor to enhance our understanding of the \textsc{Epitome} framework by conducting a comprehensive series of ablation studies. 
Our primary objective is to meticulously assess the efficacy of each individual component within the \textsc{Epitome} framework. To achieve this, we employ the x64 dataset as our focal point for analysis.

\noindent \textbf{Effect of Pretrained Assembly Language Model.}
To investigate the impact of the pre-trained model on learning function semantics, we conducted a comparison of \textsc{Epitome}'s performance with and without the utilization of pre-trained weights from the assembly language model. We evaluated and compared the performance of \textsc{Epitome} when initialized with (1) the weights of the pre-trained assembly language model, and (2) random weights.
Figure~\ref{fig:nopre} illustrates the performance of \textsc{Epitome} on the same test set. We observed that the inclusion of pre-trained weights resulted in an average precision improvement of 6.1\%. This improvement suggests that the pre-trained assembly language model effectively captures the dependencies between instructions, thereby enhancing \textsc{Epitome}'s ability to predict function names accurately.

\noindent \textbf{Effect of Multi-Task Learning.}
We also investigated how our multi-task learning framework enhances the performance of \textsc{Epitome}. We evaluated \textsc{Epitome} trained with and without the multi-task learning framework and present the results in Figure~\ref{fig:mul-task}.
The results demonstrate the effectiveness of the multi-task learning framework in improving \textsc{Epitome}'s performance. For instance, the average F1 scores of \textsc{Epitome} with and without multi-task learning are 0.67 and 0.71, respectively, indicating that multi-task learning improves \textsc{Epitome}'s performance by 5.97\%.
This finding supports our hypothesis: the distance between semantic embeddings of different optimization levels functions that come from the same source code should be close in high-dimensional vector spaces, and fusing this information into function semantics does aid in the accurate prediction of function names.

\begin{tcolorbox}
    \noindent \textbf{RQ3 Answer:} Each component of \textsc{Epitome} contributes to the overall effectiveness. For example, our proposed multi-task learning framework can improve the F1 score of \textsc{Epitome} by 5.97\% in x64 binaries.
\end{tcolorbox}

\subsection{Dataset Impact on Model}

\begin{figure}[bt]
    \centering
    \vspace{-2pt}
    \begin{minipage}{0.47\linewidth}
		\centering
		\includegraphics[width=0.9\linewidth]{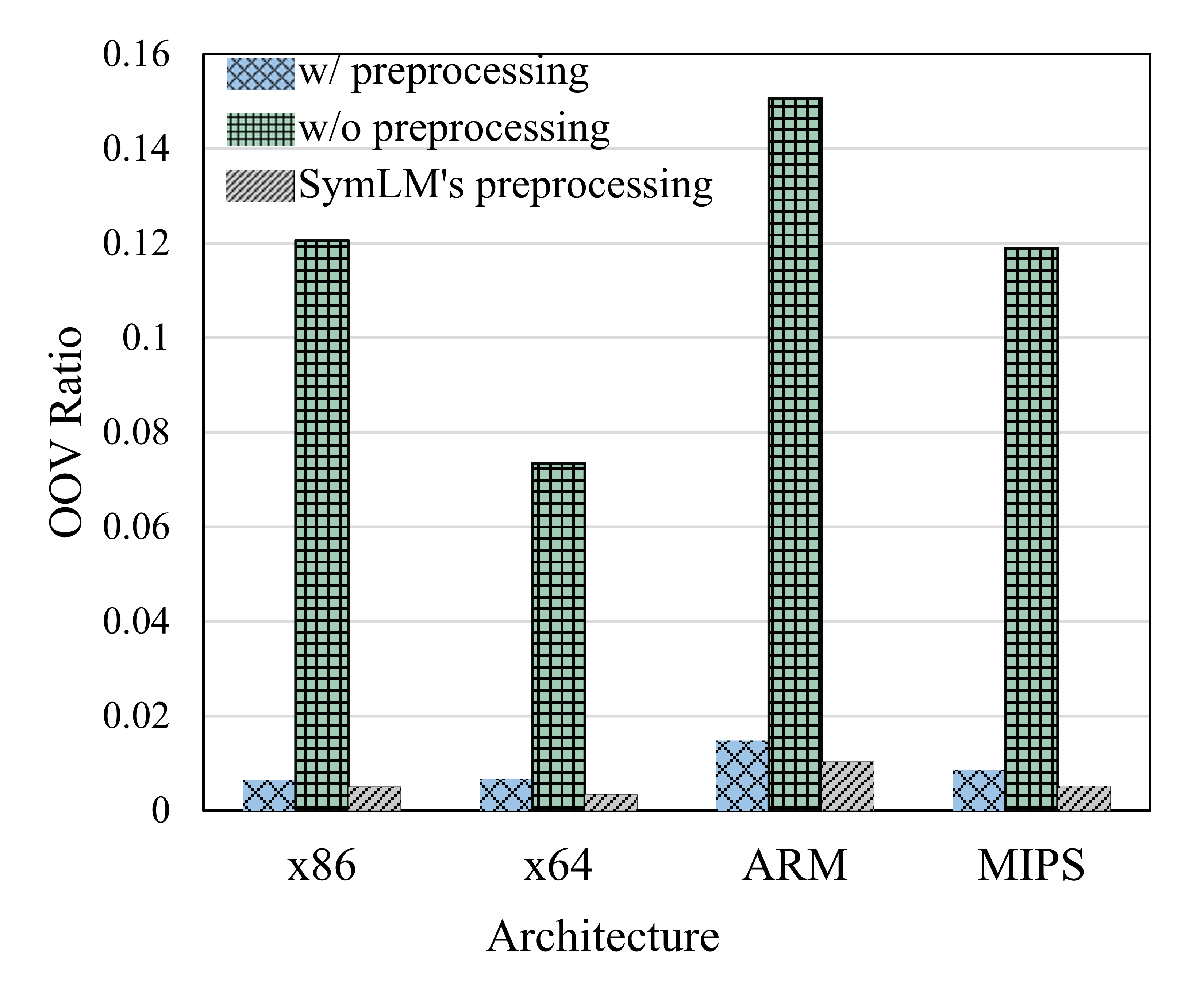}
		\caption{Effectiveness of Preprocessing on Mitigating OOV Words.}
		\label{fig:oov}
	\end{minipage}
        \vspace{-5pt}
        \begin{minipage}{0.42\linewidth}
		\centering
		\includegraphics[width=0.9\linewidth]{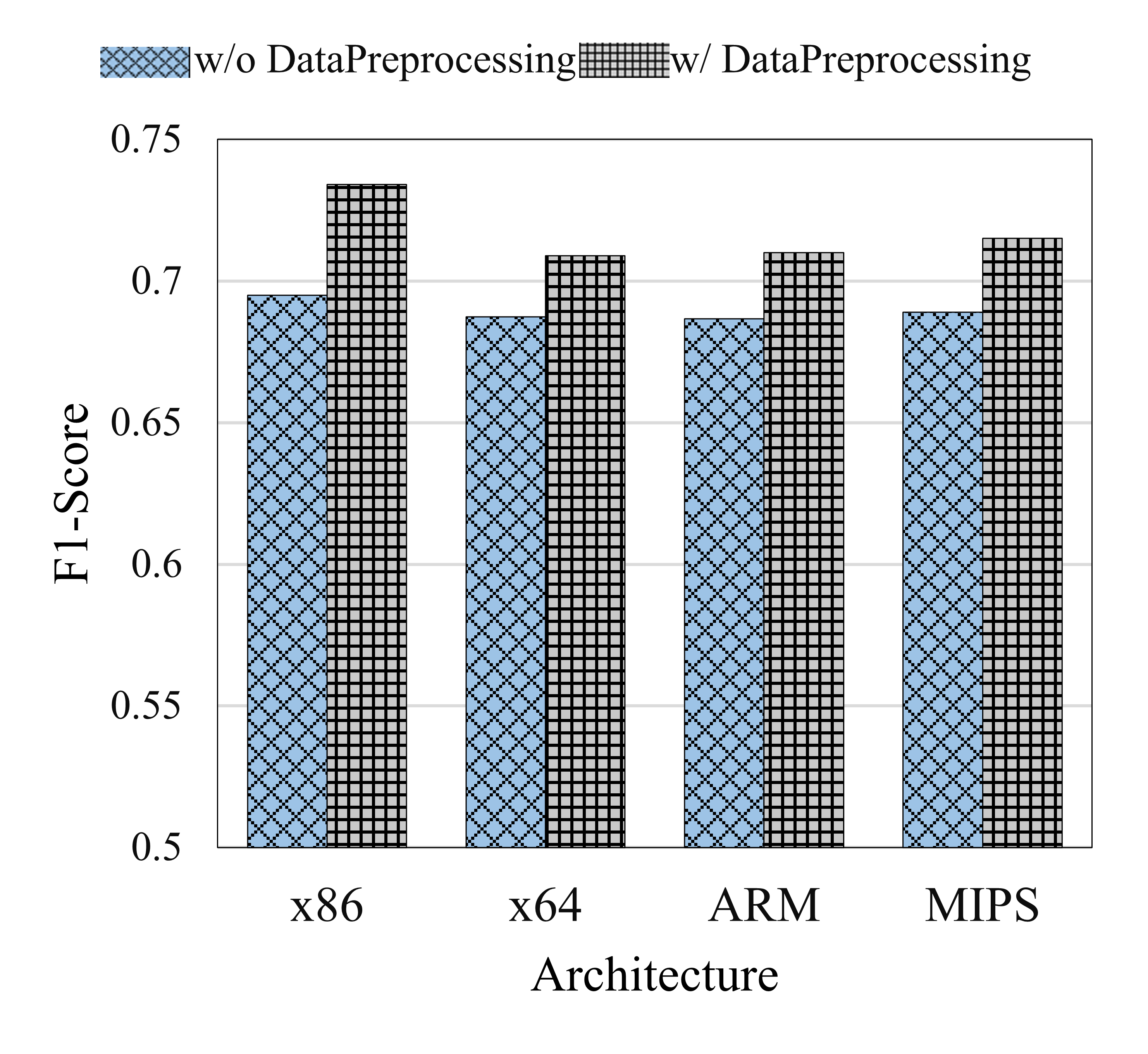}
		\caption{Effectiveness of Preprocessing on Performance Improvement.}
		\label{fig:datapact}
	\end{minipage}
    \vspace{-5pt}
\end{figure}

Based on the intuition that the issues of label sparsity and OOV will affect the performance of \textsc{Epitome}, we present data preprocessing methods described in earlier sections for mitigation. Here we validated the significance of this practice.
We conducted a comparison of the OOV word ratios obtained with and without our preprocessing approach, as well as with SymLM~\cite{b44}, on all the datasets we compiled. The OOV word ratio represents the proportion of function name labels in the test set that are not present in the vocabulary of the training sets.
We utilized the code provided by SymLM to preprocess the function names in the same datasets. For the approach without preprocessing, we tokenized function names into individual lower-case labels based on common conventions.
Figure~\ref{fig:oov} presents the OOV ratios among different architectures. 
The results demonstrate a significant reduction in the OOV word ratios when our preprocessing approach is applied. Specifically, in x86 binaries, the OOV ratio decreases from 12.06\% to 0.65\%, indicating that 94.61\% of the OOV words are effectively mitigated. Additionally, the size of the x86 dataset vocabulary is reduced from 13,004 to 6,217, effectively addressing the issue of label sparsity.

\begin{table}[bt]
\centering
    \vspace{-10pt}
    \caption{Examples of Function Name Tokenization by \textsc{Epitome} and SymLM}
    \begin{center}
    \begin{tabular}{p{3.8cm}p{3.8cm}p{3.8cm}}
    \toprule
       {\textbf{Raw Func Name}} & {\textbf{\textsc{Epitome}}} & {\textbf{SymLM}} \\
       \midrule
       test nofork sideeffects & test no fork side effect & t est no for k side eff ect s \\    
       typenameTypeMod & type name type mod & typename t y pe m od \\
       nomoreargs & no more arg s & nom ore args \\
       resolvebuiltin & resolve builtin & resolve bu il t in \\
       scanpmwidgets & scan pm widget &scan p m widget s \\
       zipfileNext & zip file next & zipfile n ext \\
    \bottomrule
    \end{tabular}
    \label{tab:oov}
    \end{center}
    \vspace{-8pt}
\end{table}

Although our proposed approach has a slightly lower performance in OOV mitigation compared to SymLM, it places greater emphasis on segmenting function names into meaningful labels, as shown in Table~\ref{tab:oov}. 
For example, the label "next" is split into \{n, ext\} by SymLM. However, the label "ext" also carries its own meaning, which is totally different from `next'.
Furthermore, the labels generated by SymLM are ordered by their confidence, different parts of the word may not be generated in order. That may mislead reverse engineers into comprehending the purpose of the function.
We believe it is justifiable to accept a minor decrease in OOV mitigation performance in order to have intelligible labels in use. 

To validate the performance improvements brought by our dataset construction methods, we conducted an evaluation of \textsc{Epitome} with and without these methods. The results, as shown in Figure~\ref{fig:datapact}, clearly demonstrate the effectiveness of our dataset construction methods. Specifically, we observed a significant improvement in \textsc{Epitome}'s F1 score by 5.6\% on the x86 binaries.

\begin{tcolorbox}
    \noindent \textbf{RQ4 Answer:} The issues of label sparsity and OOV indeed hurt the performance of the model. \textsc{Epitome} trained on the raw dataset has a lower performance than that on the preprocessed dataset. It is significant to perform data preprocessing for mitigation.
\end{tcolorbox}

\section{Illustration}
\subsection{Qualitative Evaluation}
In our qualitative analysis of the predicted function names by \textsc{Epitome}, we found interesting insights. Table~\ref{tab:case_study1} showcases examples of predictions, including the ground-truth function names before and after preprocessing, as well as the predicted function name words.
We identified four categories of prediction names, as outlined in Table~\ref{tab:case_study1}:
(i) Our models accurately predict all the ground truth words.
(ii) Our predictions have semantically related meanings to the ground truth, although different words are used (e.g., \{find, match\}).
(iii) Our predictions include additional words not present in the ground truth (e.g., the predicted word "recover" alongside \{log, sign\}).
(iv) Our predictions miss some ground truth words, such as "recover" for \{proxy, sign, recover\}.
Interestingly, we observed that despite some erroneous predictions, certain predictions captured the functional semantics of the ground truth, even if not reflected in the metrics. For example, the F1 score of our prediction of "attrs find" is 0.5, but it has the same meaning as its ground truth.
This implies that the actual performance of \textsc{Epitome} might surpass the reported number.

\begin{table}[bt]
    \caption{Qualitative Examples of Predictions Made by \textsc{Epitome}}
    \vspace{-8pt}
    \centering
    \begin{tabular}{p{4.1cm}p{3.5cm}p{3.8cm}}
    \toprule
    \multicolumn{2}{c}{\textbf{Ground Truth}}&\textbf{Prediction}\\
    \cline{0-1}
    \textbf{Raw Label} & \textbf{Preprocessed} &  \\
    \midrule
         atexit& at exit & at exit\\
         getPasswd12 & get pass wd & get pass wd \\
         \hline
         p11\_attrs\_match & attrs match & attrs find \\
         base\_C\_DigestUpdate &base digest update & base sign update \\
         \hline
         binding\_C\_Decrypt\_Update & bind decrypt update & bind decrypt verify update \\
         log\_C\_Sign & log sign & log sign recover \\
         \hline
         ErrorExitWarn& error exit warn & exit \\
         proxy\_C\_SignRecover & proxy sign recover & proxy sign \\
         \bottomrule
    \end{tabular}
    \label{tab:case_study1}
    \vspace{-5pt}
\end{table}

\subsection{Large Language Model Comparison}
\begin{table*}[bt]
    \centering
    \caption{Comparison Between \textsc{Epitome} and ChatGPT}
    \resizebox{\columnwidth}{!}{
    \begin{tabular}{ccc}
        \toprule
         \textbf{Raw Label} & \textbf{\textsc{Epitome}} & \textbf{ChatGPT}  \\
         \midrule
         rpc\_generate\_key\_pair&rpc\_generate\_key\_pair&prepare\_and\_send\_rpc\_message \\
         binding\_C\_SetAttributeValue& bind\_set\_attribute\_value  & invoke\_function\_from\_table \\
         proxy\_C\_GetTokenInfo & proxy\_get\_token\_information& map\_slot\_and\_invoke\_function\\
         p11\_kit\_registered\_name\_to\_module &kit\_register\_name\_to\_module&find\_module\_by\_name\_thread\_safe\\
         mock\_C\_InitToken\_specific\_args&mock\_initialize\_token\_specific\_args&handle\_pin\_and\_label\_validation \\
         \bottomrule
    \end{tabular}
    }
    \label{tab:chatgpt}
    \vspace{-10pt}
\end{table*}
Large language model ChatGPT~\cite{chatgpt} is a natural language processing tool driven by existing advanced artificial intelligence technology. 
In order to verify the effectiveness of \textsc{Epitome}, we sampled some binary functions to generate names by querying ChatGPT. The used prompt is proposed by LmPa~\cite{xu2023lmpa}. 
Some examples are shown in Table~\ref{tab:chatgpt}. Since we preprocess the function name, the names generated by \textsc{Epitome} cannot exactly match the ground truth. 
Compared with the query results of ChatGPT, experimental results show that \textsc{Epitome} can better rename binary functions. ChatGPT can generate partial ground truth, but it fails to capture key semantic information of function. 
This may be due to the fact that the main application of ChatGPT is natural language processing, which lacks domain knowledge related to binary analysis during training.
Our experiments reveal that \textsc{Epitome} outperforms ChatGPT in renaming binary functions. This study underscores the ongoing significance of function name prediction research for binaries, even in the context of emerging large language models.

\subsection{Firmware Analysis}
To gain insights into practical applications, we utilized \textsc{Epitome} to annotate functions in firmware samples, drawing upon open-source firmware datasets from previous analysis studies~\cite{sfuzz, P2IM}. We specifically focused on eight 32-bit ARM firmware images, which had also been examined in SymLM's firmware evaluation.
From these firmware binaries, we extracted 1,061 functions and then preprocessed the ground truth using our proposed approaches. This preprocessing revealed that 18.39\% of the function name labels were not present in our ARM dataset vocabulary, indicating a significant OOV ratio. This high OOV highlights the unique lexicon of IoT device firmware, featuring terms like BLPE, NVIC, and HAL, distinct from the vocabulary used in \textsc{Epitome}'s training dataset. In predicting function names within these firmware samples using \textsc{Epitome}, trained on our ARM dataset, we achieved partially accurate predictions for 242 functions.
For example, for the ground truth function name \{\_\_aeabi\_cdrcmple\}, our prediction was  \{aeabi, cdr, compare, le\}. Despite the absence of 'ul' in our prediction for  \{ \_\_aeabi\_uldivmod\}, the predicted \{aeabi, abi, div, mod\} retains the fundamental essence of the original function name.

Additionally, we extended our analysis to firmware based on VxWorks~\cite{vxworks}, using the 32-bit ARM firmware TPlink\_wdr7660 provided by SFuzz~\cite{sfuzz}, as a case study. From this firmware, we extracted 7,168 functions and applied our preprocessing techniques to the ground truth.
This process revealed a 24.8\% OOV ratio, likely due to its IoT-VxWorks nature diverging from our Linux-based ARM training data.
Applying \textsc{Epitome}, trained on our ARM data, achieved partial accuracy for 558 functions. The manual inspection found many `incorrect' predictions to be semantically similar to their true names (e.g., the prediction `rsa function' is semantically akin to `mcrypt decrypt'), suggesting that \textsc{Epitome}'s practical utility transcends standard precision measures.
Notably, the KL divergence between our ARM training dataset and the TPlink\_wdr7660 firmware was $1.2\times$ higher compared to divergence with SymLM’s firmware, highlighting distribution shifts that hinder \textsc{Epitome}'s efficiency.
Further, training \textsc{Epitome} on a VxWorks dataset and applying it to TPlink\_wdr7660 yielded partial correctness for 2,085 functions, supporting our earlier observation in (\ref{generalizability}): \textsc{Epitome} excels with unfamiliar binaries sharing domain-specific similarities to training data.

\section{DISCUSSION \& LIMITATION}
This section discusses the limitations of our work and future directions. 

\noindent \textbf{Method of Information Extract.}
Our primary focus in this study is the information extracted through static analysis, and we do not incorporate information obtained from dynamic execution. Dynamic analysis can indeed offer a more comprehensive understanding of binary behavior, but its execution can be time-consuming and dependent on specific environments. 
It requires the trade-offs between the increased computational cost and the improved prediction accuracy.

\noindent \textbf{Noise in Ground Truth.}
Although we have made efforts to address different naming methods, it is important to acknowledge that the rule-based approach used in our dataset construction method may still have limitations in fully solving the problem. Rule-based approaches rely on predefined rules to tokenize function names, which may not capture all uncommon naming practices.

\noindent \textbf{Label Imbalance in Function Names.}
We have observed that the labels in the dataset exhibit a long-tailed distribution, where a few labels occur frequently while the majority of labels appear infrequently. This distribution presents a challenge for the model, as it tends to prioritize high-frequency labels and may overlook the learning of low-frequency labels. Addressing the long-tail distribution of data is a crucial and promising area of focus in machine learning-based applications.

\noindent \textbf{Hyperparameter Selection.}
Hyperparameters are pivotal for model performance, being manually predefined. In the absence of a systematic tuning process, we resorted to empirical methods for setting \textsc{Epitome}'s hyperparameters. The compared method adheres to its published hyperparameter configuration, albeit with no assurance of reproducing the documented performance. A common safeguard adopted across methods is the use of an early stopping strategy to curb overfitting.

\section{RELATED WORK}
In this section, we briefly survey the related work about function name prediction.

\subsection{Predict Function Names From Source Code} 
Predicting function names from source code is commonly known as source code summarization, which involves generating readable natural language descriptions for code snippets. Since around 2016, this field has predominantly seen data-driven methodologies, with Iyer et al. pioneering the use of attention-based encoder-decoders akin to machine translation~\cite{b15}. Building upon this approach, ~\citet{b16} utilized two encoders, namely an API encoder and a code encoder, to capture the semantics of the source code. Subsequent advancements aimed at extracting more information from the code.
~\citet{b17} and ~\citet{b18}  flattened the Abstract Syntax Tree (AST) into a sequence for encoder-decoder inputs. \citet{b20} and \citet{b21} employed graph-based neural networks to encode the AST and generate natural language summaries.
All of the methods primarily rely on the rich textual information within source code functions and incorporate structured information as auxiliary data to enhance the completion of source code summarization tasks.

\subsection{Predict Function Names From Binary Code}
Debin~\cite{b5} and Punstrip~\cite{b7} are notable works in predicting function names from stripped binaries using machine learning. 
Debin transforms binary code into an intermediate form and builds a variable dependency graph, employing a Conditional Random Field (CRF) model for predictions. Meanwhile, Punstrip creates probabilistic binary fingerprints from manually designed features, using a CRF-based methodology to infer connections between function names and program structure.
In contrast, Nero~\cite{b4} models function name prediction as neural machine translation, focusing on call contexts through graph construction but overlooking non-call instructions.
While these methods have made advancements in function name prediction, they do have limitations. 
The utilization of CRF in Debin and Punstrip makes them difficult to generalize to new names, as they only assign function names from a pre-determined closed set. 
Nero heavily relies on semantic information from library function names, which may not be available for functions without such references.
It's worth mentioning that NFRE~\cite{b6} and SymLM~\cite{b44} are discussed in~\ref{relate_work} and not be introduced here.

\section{CONCLUSION}
We introduce \textsc{Epitome}, a novel neural architecture for predicting binary function names using a multi-task learning framework. 
\textsc{Epitome} includes an auxiliary task that predicts the similarity of function semantics, with the goal of maximizing the similarity between function semantics with the same name but different optimization levels.
To capture the underlying characteristics of instructions, we utilize a pre-trained assembly language model to generate general-purpose instruction embeddings. 
We also design a novel semantics encoder module to effectively model the sequence and structural information of functions.
Furthermore, we propose two data-preprocessing approaches to address the issues of label sparsity and OOV. 
Through our evaluation, we demonstrate that \textsc{Epitome} outperforms state-of-the-art tools by up to 54.44\% in terms of F1 score. We also showcase the effectiveness of its components and highlight its potential usability.

\textbf{Data Availability:} To promote open science, we have made our code publicly available at the following link: \url{https://github.com/Xiaolinger-Z/Epitome}.

\begin{acks}
We thank the associated editors and reviewers of FSE 2024 for their valuable feedback. This research was supported in part by
the Beijing Natural Science Foundation (Grant No. L234033) and 
Joint Fund Cultivation Project of National Natural Science Foundation of China (Grant No. U1636120).
\end{acks}
\clearpage
\bibliographystyle{ACM-Reference-Format}
\bibliography{references}


\end{document}